\documentclass{aa}

\usepackage[varg]{txfonts}
\usepackage{graphicx}
\usepackage{color} 
\usepackage{amsmath}
\usepackage{multirow} 
\bibpunct{(}{)}{;}{a}{}{,} 

\begin{document}

\title{Robustness of N$_2$H$^+$ as tracer of the CO snowline}

\author{M.L.R. van 't Hoff\inst{1}
\and C. Walsh\inst{1,2} 
\and M. Kama\inst{1,3} 
\and S. Facchini\inst{4}
\and E.F. van Dishoeck\inst{1,4}}

\institute{Leiden Observatory, Leiden University, P.O. box 9513, 2300 RA Leiden, The Netherlands
\and School of Physics and Astronomy, University of Leeds, Leeds, UK, LS2 9JT
\and Institute of Astronomy, Madingley Road, Cambridge, UK, CB3 0HA
\and Max-Planck-Institut f\"ur Extraterrestrische Physik, Giessenbachstrasse 1, 85748 Garching, Germany }

\date{}

\abstract {Snowlines in protoplanetary disks play an important role in planet formation and composition. Since the CO snowline is difficult to observe directly with CO emission, its location has been inferred in several disks from spatially resolved ALMA observations of DCO$^+$ and N$_2$H$^+$.} 
{N$_2$H$^+$ is considered to be a good tracer of the CO snowline based on astrochemical considerations predicting an anti-correlation between N$_2$H$^+$ and gas-phase CO. In this work, the robustness of N$_2$H$^+$ as a tracer of the CO snowline is investigated.} 
{A simple chemical network is used in combination with the radiative transfer code LIME to model the N$_2$H$^+$ distribution and corresponding emission in the disk around TW Hya. The assumed CO and N$_2$ abundances, corresponding binding energies, cosmic ray ionization rate, and degree of large-grain settling are varied to determine the effects on the N$_2$H$^+$ emission and its relation to the CO snowline.}
{For the adopted physical structure of the TW Hya disk and molecular binding energies for pure ices, the balance between freeze-out and thermal desorption predicts a CO snowline at 19 AU, corresponding to a CO midplane freeze-out temperature of 20 K. The N$_2$H$^+$ column density, however, peaks 5--30~AU outside the snowline for all conditions tested. In addition to the expected N$_2$H$^+$ layer just below the CO snow surface, models with an N$_2$/CO ratio $\gtrsim$ 0.2 predict an N$_2$H$^+$ layer higher up in the disk due to a slightly lower photodissociation rate for N$_2$ as compared to CO. The influence of this N$_2$H$^+$ surface layer on the position of the emission peak depends on the total CO and N$_2$ abundances and the disk physical structure, but the emission peak generally does not trace the column density peak. A model with a total, i.e. gas plus ice, CO abundance of $3 \times 10^{-6}$ with respect to H$_2$ fits the position of the emission peak observed by \citet{Qi2013} for the TW Hya disk.}
{The relationship between N$_2$H$^+$ and the CO snowline is more complicated than generally assumed: for the investigated parameters, the N$_2$H$^+$ column density peaks at least 5~AU outside the CO snowline. Moreover, the N$_2$H$^+$ emission can peak much further out, as far as $\sim$50~AU beyond the snowline. Hence, chemical modeling, as done here, is necessary to derive a CO snowline location from N$_2$H$^+$ observations.}


\maketitle


\section{Introduction}

Protoplanetary disks around young stars contain the gas and dust from which planetary systems will form. In the midplanes of these disks, the temperature becomes so low that molecules freeze out from the gas phase onto dust grains. The radius at which this happens for a certain molecule is defined as its snowline. The position of a snowline depends both on the species-dependent sublimation temperature and disk properties (mass, temperature, pressure and dynamics). Snowlines play an important role in planet formation as increased particle size, surface density of solid material, and grain stickiness at a snowline location may enhance the efficiency of planetesimal formation \citep{Stevenson1988,Ciesla2006,Johansen2007,Chiang2010,Gundlach2011,Ros2013}. Furthermore, the bulk composition of planets may be regulated by the location of planet formation with respect to snowlines, as gas composition and ice reservoirs change across a snowline \citep{Oberg2011,Madhusudhan2014,Walsh2015,Eistrup2016}. Determining snowline locations is thus key to studying planet formation. 

\begin{figure*}
\centering
\includegraphics[width=17cm,trim={0 16cm .5cm 1cm},clip]{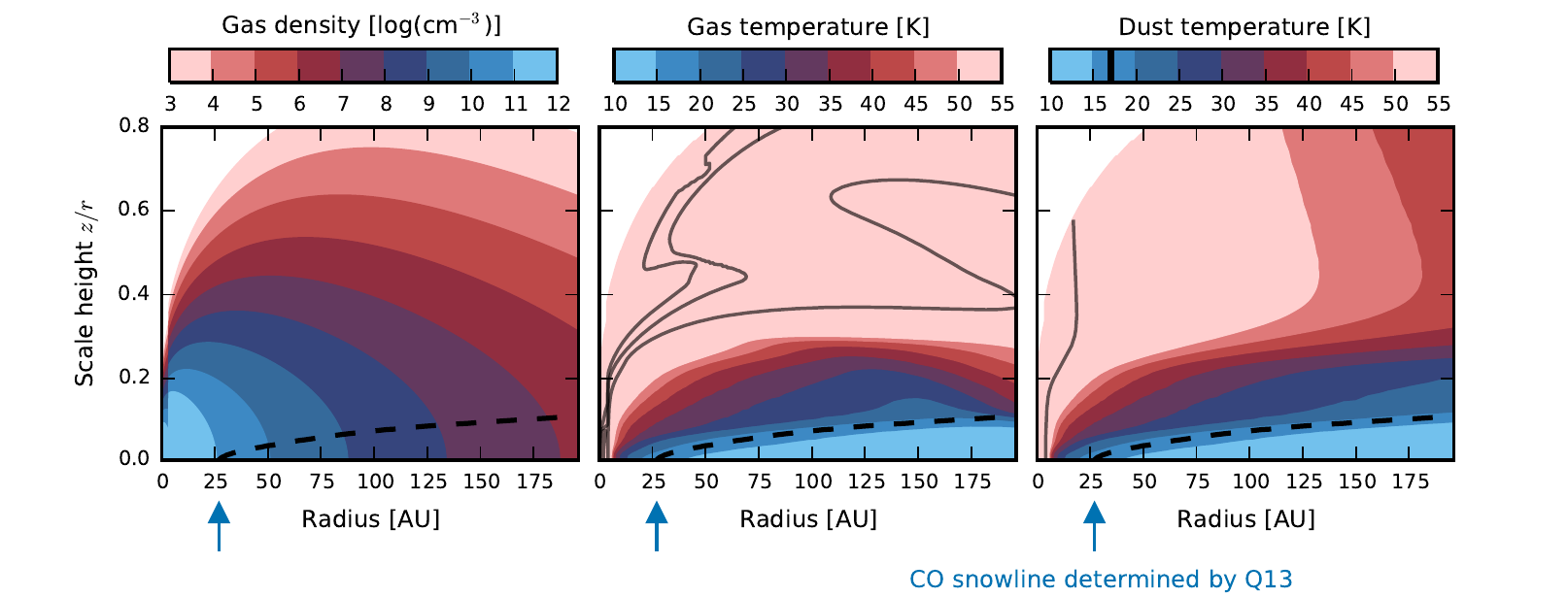}
\caption{Gas density (cm$^{-3}$), gas temperature (K), and dust temperature (K) as a function of disk radius, \textit{r}, and scale height, \textit{z/r}, for the adopted model for the TW Hya disk. The temperature color range is limited to highlight values around the CO snow surface. The solid black contours indicate temperatures of 100, 200 and 500 K. The blue arrow indicates the location of the midplane CO snowline associated with a freeze-out temperature of 17 K, as determined by Q13, and the dashed contour marks the corresponding snow surface.}
\label{fig:PhysicalStructure}
\end{figure*}

The CO snowline is of particular interest because CO ice is a starting point for prebiotic chemistry \citep{Herbst2009}. Assuming a disk around a solar-type star, the CO snowline occurs relatively far (a few tens of AU) from the central star due to the low freeze-out temperature of CO; hence, it is more accessible to direct observations than other snowlines. However, locating it is difficult because CO line emission is generally optically thick, so that the bulk of the emission originates in the warm surface layers. An alternative approach is to observe molecules whose emission is expected to peak around the snowline, or molecules that are abundant only when CO is depleted from the gas phase. Based on the former argument, DCO$^+$ has been used to constrain the CO snowline location \citep{Mathews2013,Oberg2015}, but may be affected by some DCO$^+$ also formed in warm disk layers \citep{Favre2015,Qi2015}. A species from the latter category is N$_2$H$^+$ \citep{Qi2013,Qi2015}. This molecule forms through proton transfer from H$_3^+$ to N$_2$, 
\begin{equation} \label{eq:N2H+formation}
\mathrm{N_2 + H_3^+} \rightarrow \mathrm{N_2H^+ + H_2},
\end{equation} 
but provided that CO is present in the gas phase, its formation is impeded, because CO competes with N$_2$ for reaction with H$_3^+$,  
\begin{equation} \label{eq:HCO+formation}
\mathrm{CO + H_3^+} \rightarrow \mathrm{HCO^+ + H_2}. 
\end{equation} 
Furthermore, reactions with CO are the dominant destruction pathway of N$_2$H$^+$: 
\begin{equation} \label{eq:N2H+destruction}
  \mathrm{N_2H^+ + CO} \rightarrow \mathrm{HCO^+ + N_2}.
\end{equation}
N$_2$H$^+$ is therefore expected to be abundant only in regions where CO is depleted from the gas phase, i.e., beyond the CO snowline. 

Observational evidence for the anti-correlation of N$_2$H$^+$ and gas-phase CO was initially provided for pre-stellar and protostellar environments \citep[e.g.][]{Caselli1999,Bergin2001,Jorgensen2004b}. However, survival of N$_2$H$^+$ is aided in these systems by the delayed freeze-out of N$_2$ as compared to CO, because gas-phase N$_2$ forms more slowly when starting from atomic abundances under diffuse cloud conditions \citep{Aikawa2001,Maret2006}. In protoplanetary disks, N$_2$ molecules are expected to be more abundant than N atoms because of the higher gas density which increases the N$_2$ formation rate, and this timescale effect is not important. 

So far, the results for protoplanetary disks seem inconclusive. Recent observations of C\element[][18]{O} in the disk of HD 163296 suggest a CO snowline location consistent with the observed N$_2$H$^+$ emission \citep{Qi2015}. On the other hand, several studies indicate a depletion of CO in the disk around TW Hya down to $\sim$10~AU \citep{Favre2013,Nomura2016,Kama2016,Schwarz2016}, inconsistent with the prediction that CO is depleted only beyond a snowline at $\sim$30~AU, based on modeling of N$_2$H$^+$ observations \citep[][hereafter Q13]{Qi2013}.

In this work, we explore the robustness of the N$_2$H$^+$ line emission as a tracer of the CO snowline location in the disk midplane, using a physical model (constrained by observations) for the disk around TW Hya. TW Hya is the closest protoplanetary disk system \citep[$\sim$ 54 pc,][]{vanLeeuwen2007} and considered an analog of the Solar Nebula based on disk mass and size. The spatial distribution and emission of N$_2$H$^+$ are modeled for different CO and N$_2$ abundances and binding energies, as well as different cosmic ray ionization rates and degrees of dust settling, using a simple chemical network and full radiative transfer. \citet{Aikawa2015} have shown that analytical formulae for the molecular abundances give a similar N$_2$H$^+$ distribution as a full chemical network. They also found that the N$_2$H$^+$ abundance can peak at temperatures slightly below the CO freeze-out temperature in a typical disk around a T Tauri star, but they did not invoke radiative transfer to make a prediction for the resulting N$_2$H$^+$ emission.  

The physical and chemical models used in this work are described in Sect.~\ref{sec:Models}, and Sect.~\ref{sec:Results} shows the predicted N$_2$H$^+$ distributions and emission. The simulated emission is compared with that observed by Q13 and convolved with a smaller beam ($0\farcs2\times0\farcs2$) to predict results for future higher angular resolution observations. This section also studies the dependence of the model outcome on CO and N$_2$ abundances, binding energies, cosmic ray ionization rate, and dust grain settling, and the use of multiple N$_2$H$^+$ transitions to further constrain the snowline location. Finally, the dependence of the outer edge of the N$_2$H$^+$ emission on chemical and physical effects is explored. In Sect.~\ref{sec:Discussion} the implications of the results will be discussed and in Sect.~\ref{sec:Conclusions} the conclusions summarized.


\section{Protoplanetary disk model} \label{sec:Models}


\subsection{Physical model} \label{sec:Physicalmodel}

For the physical structure we adopt the model for TW Hya from \citet{Kama2016}. This model reproduces the dust spectral energy distribution (SED) as well as CO rotational line profiles, from both single-dish and ALMA observations, and spatially resolved CO $J$=3--2 emission from ALMA. The 2D physical-chemical code DALI \citep[Dust And LInes,][]{Bruderer2009,Bruderer2012,Bruderer2013} was used to create the model, assuming a stellar mass and radius of \mbox{$M_*$ = 0.74 $\mathrm{M}_{\sun}$} and \mbox{$R_*$ = 1.05 $\mathrm{R}_{\sun}$}, respectively. The disk is irradiated by UV photons and X-rays from the central star and UV photons from the interstellar radiation field. The stellar UV spectrum from \citet{Cleeves2015}  is used (based on \citealt{Herczeg2002,Herczeg2004} and \citealt{France2014}), which is roughly consistent with a $\sim$~4000~K blackbody with UV excess due to accretion. The X-ray spectrum is modeled as a thermal spectrum at \mbox{$3.2 \times 10^6$ K} with a total X-ray luminosity of \mbox{$1.4 \times 10^{30}$ erg s$^{-1}$} and the cosmic ray ionization rate is taken to be low, 5 $\times$ 10$^{-19}$ s$^{-1}$ \citep{Cleeves2015}. 

Starting from an input gas and dust density structure the code uses radiative transfer to determine the dust temperature and local radiation field. The chemical composition is obtained from a chemical network simulation based on a subset of the UMIST 2006 gas-phase network \citep{Woodall2007} and used in a non-LTE excitation calculation for the heating and cooling rates to derive the gas temperature (see \citealt{Bruderer2012} for details). As will be shown in Sect.~\ref{sec:Results} and Fig.~\ref{fig:PhysicalStructure}, N$_2$H$^+$ is predicted in the region where the gas and dust temperatures are coupled ($z/r \lesssim 0.25$). Hence, the temperature in the relevant disk region is not sensitive to changes in molecular abundances.  

The input gas density has a radial power law distribution, 
\begin{equation}
\Sigma_{\mathrm{gas}} = 30 \, \mathrm{g \, cm}^{-2} \left( \frac{r}{35 \, \mathrm{ AU}} \right) ^{-1} \exp \left( \frac{-r}{35 \, \mathrm{ AU}} \right),
\end{equation}
and a vertical Gaussian distribution, 
\begin{equation}
h = 0.1 \left( \frac{r}{35 \, \mathrm{ AU}} \right) ^{0.3}. 
\end{equation}
To match the observations, the gas-to-dust mass ratio is set to 200. Two different dust populations are considered; small grains (0.005 - 1 $\mu$m) represent 1\% of the dust surface density, whereas the bulk of the dust surface density is composed of large grains (0.005 - 1000 $\mu$m). The vertical distribution of the dust is such that large grains are settled toward the midplane with a settling parameter $\chi$ of 0.2, i.e. extending to 20\% of the scale height of the small grains;
\begin{equation}
  \rho_{\mathrm{dust,small}} = \frac{0.01\Sigma_{\mathrm{dust}}}{\sqrt{2\pi}Rh} \exp \left[ -\frac{1}{2} \left( \frac{\pi/2-\theta}{h} \right) ^2 \right] \hspace{0.2cm} \mathrm{g \, cm}^{-3}, \mathrm{and}
\end{equation}
\begin{equation}
  \rho_{\mathrm{dust,large}} = \frac{0.99\Sigma_{\mathrm{dust}}}{\sqrt{2\pi}R\chi h} \exp \left[ -\frac{1}{2} \left( \frac{\pi/2-\theta}{\chi h} \right) ^2 \right] \hspace{0.2cm} \mathrm{g \, cm}^{-3},
\end{equation}
\\where $\theta$ is the vertical latitude coordinate measured from the pole ($\theta = 0$) to the equator, i.e. the midplane ($\theta = \pi/2$; \citealt{Andrews2012}). In the inner 4 AU, the gas and dust surface density is lowered by a factor of 100 with respect to the outer disk to represent the gap detected in the inner disk \citep{Calvet2002,Hughes2007}. Recent observations indicate that the dust distribution in this inner region is more complicated \citep{Andrews2016}, but this will not affect the N$_2$H$^+$ distribution in the outer disk. In Sect.~\ref{sec:GrainSettling} we examine the influence of grain settling on the N$_2$H$^+$ distribution and emission by using a model with $\chi$ = 0.8, i.e. the large grains extending to 80\% of the small grain scale height.

The resulting density and thermal structure of the disk are shown in Fig.~\ref{fig:PhysicalStructure} and used in the chemical modeling described in Sect.~\ref{sec:Chemicalmodel}. A midplane temperature of 17 K corresponds to a radius of 27.5 AU, consistent with the CO snowline properties derived by Q13. In their analysis, Q13 fit ALMA observations using a power law for the radial distribution of the N$_2$H$^+$ column density, with an inner radius presumed to coincide with the CO snowline. 


\begin{figure}[!t]
\resizebox{\hsize}{!}
{\includegraphics[trim={0 1cm 0cm .4cm},clip]{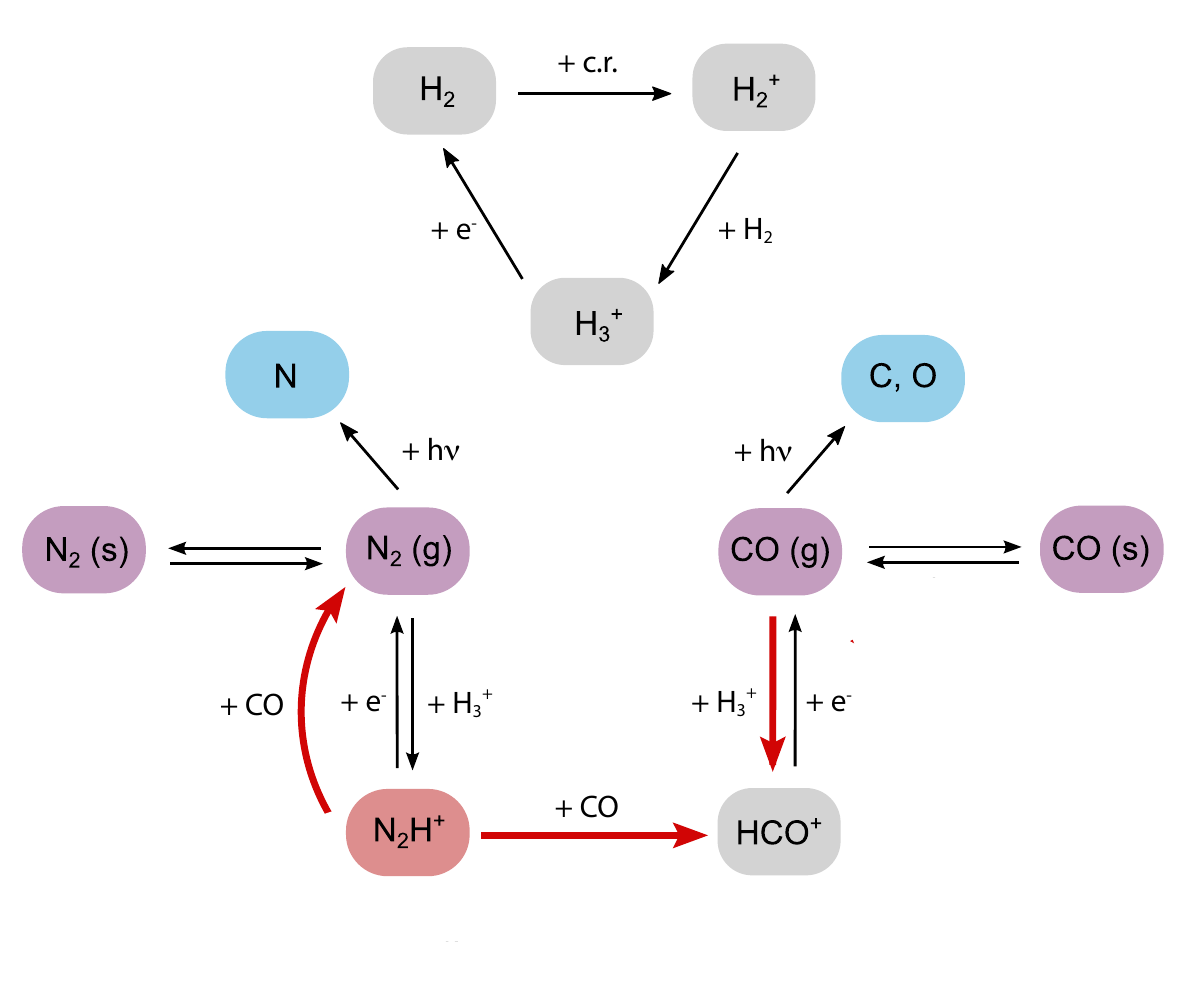}}
\caption{Schematic representation of the chemical network used to model N$_2$H$^+$ (red). Freeze-out and desorption products are highlighted in purple and photodissociation products are shown in blue. The processes responsible for the anti-correlation between N$_2$H$^+$ and CO are highlighted with red arrows. }
\label{fig:ChemicalNetwork}
\end{figure}

\subsection{Chemical model} \label{sec:Chemicalmodel}

If CO is abundant in the gas phase, N$_2$H$^+$ formation is slowed down (Eqs.~\ref{eq:N2H+formation}~and~\ref{eq:HCO+formation}) and N$_2$H$^+$ destruction is enhanced (Eq.~\ref{eq:N2H+destruction}). On the other hand, gas-phase N$_2$ is required to form N$_2$H$^+$ (Eq.~\ref{eq:N2H+formation}). Based on these considerations, the simplest method to predict the distribution of N$_2$H$^+$ is by calculating the balance between freeze-out and desorption for N$_2$ and CO at every position in the disk. Assuming a constant total abundance, i.e. $n_\mathrm{g}$(CO) + $n_\mathrm{s}$(CO) = $n$(CO), the steady state gas phase and ice abundances ($n_\mathrm{g}$ and $n_\mathrm{s}$, resp.) are then given by,
\begin{eqnarray}
  \label{eq:n_gas}
  n_\mathrm{g}(\mathrm{CO}) &=& \frac{n(\mathrm{CO})}{k_\mathrm{f}/k_\mathrm{d} + 1} \hspace{0.1cm} \mathrm{cm}^{-3}, \mathrm{and}
  \\n_\mathrm{s}(\mathrm{CO}) &=& n(\mathrm{CO}) - n_\mathrm{g}(\mathrm{X}) \hspace{0.1cm} \mathrm{cm}^{-3},
  \label{eq:n_ice}
\end{eqnarray} 
where $k_\mathrm{f}$ and $k_\mathrm{d}$ are the freeze-out and desorption rates, respectively. For N$_2$ a similar equation holds. Thermal desorption is considered here as the only desorption process, which is appropriate for volatile molecules such as CO and N$_2$. However, the dust density in the outer disk may be low enough for UV photons to penetrate to the disk midplane, such that photodesorption may become effective. Photodesorption is therefore included when studying the outer edge of the N$_2$H$^+$ emission in Sect.~\ref{sec:OuterEdge}. The thermal desorption rate depends on the specific binding energy for each molecule, $E_b$, and for CO and N$_2$ values of 855 and 800 K \citep{Bisschop2006} are adopted, respectively. Expressions for the freeze-out and desorption rates, and a discussion of the adopted parameters can be found in Appendix~\ref{ap:chemmodel}. Solving the gas and ice abundances time dependently shows that equilibrium is reached within $10^5$ years, so steady state is a reasonable assumption for a typical disk lifetime of $10^6$ years.

The snow surface is defined as the position in the disk where 50\% of a species is present in the gas phase and 50\% is frozen onto the grains. From Eq.~\ref{eq:n_gas} the snow surfaces for CO and N$_2$ can thus be predicted. Note that the freeze-out and desorption rates (Eqs. \ref{eq:k_freezeout} and  \ref{eq:k_desorption}), and therefore the fraction of a species that is present in the gas or ice (e.g. $n_g$(CO)/$n$(CO); see Eq.~\ref{eq:n_gas}) at a certain temperature, do not depend on abundance. Hence the locations of the midplane snowlines are independent of the total, i.e. gas plus ice, CO and N$_2$ abundances.

As a first approximation, N$_2$H$^+$ can be considered to be present between the CO and N$_2$ snow surfaces. Comparison with the result from the chemical model described below shows that the N$_2$H$^+$ layer extends beyond the N$_2$ snow surface, and the outer boundary is better described by the contour where only 0.05\% of the N$_2$ has desorbed while the bulk remains frozen out. We will refer to the N$_2$H$^+$ layer bounded by the CO snow surface and the contour where 0.05\% of the N$_2$ has desorbed as model ``FD'' (Freeze-out and Desorption). 

\begin{table*}
\caption{Reactions, rate data and related parameters for the N$_2$H$^+$ chemical network. 
\label{tab:RateCoefficients}} 
\centering
\begin{tabular}{l c c c c c c c c }
    \hline\hline
    \\[-.3cm]
    Reaction & $\zeta$\tablefootmark{a} & $\alpha$\tablefootmark{b} & $\beta$\tablefootmark{b} & $\gamma$\tablefootmark{b} & $S$\tablefootmark{c} & $E_\mathrm{b}$\tablefootmark{d} & $Y$\tablefootmark{e} & $k_0(r,z)$\tablefootmark{f}   \\
                   & s$^{-1}$ & cm$^3$ s$^{-1}$ & & K & & K & photon$^{-1}$ & s$^{-1}$ \\
    \hline 
    \\[-.3cm]
    H$_2$ + cosmic ray $\rightarrow$ H$_2^+$ + e$^-$ & $1.20\times10^{-17}$ & ... & ...& ... & ...& ...& ...& ...\\
    H$_2^+$ + H$_2 \rightarrow$ H$_3^+$ + H & ... & $2.08\times10^{-9}$ & 0  & 0  & ... & ... & ... & ... \\
    H$_3^+$ + e$^- \rightarrow$ H$_2$ + H & ... & $2.34\times10^{-8}$ & -0.52 & 0  & ... & ... & ... & ... \\
    N$_2$ + H$_3^+ \rightarrow$ N$_2$H$^+$ + H$_2$ & ... & $1.80\times10^{-9}$ & 0 & 0  & ...&...&... & ...\\
    CO + H$_3^+ \rightarrow$ HCO$^+$ + H$_2$ & ... & $1.36\times10^{-9}$ & -0.14 & -3.4 & ... & ...&... & ...\\
    N$_2$H$^+$ + CO $\rightarrow$ HCO$^+$ + N$_2$ & ... & \hspace{0.1cm}$8.80\times10^{-10}$ & 0 & 0 & ... & ...&... &...\\
    HCO$^+$ + e$^- \rightarrow$ CO + H  & ... & $2.40\times10^{-7}$ & -0.69 & 0   & ... & ... & ... & ...  \\
    N$_2$H$^+$ + e$^- \rightarrow$ N$_2$ + H & ...  & $2.77\times10^{-7}$ & -0.74 & 0  & ... & ... & ... & ...\\
    CO $\rightarrow$ CO (ice)   & ... & ... & ...  & ...  & 0.90 & ... & ... & ... \\
    N$_2$ $\rightarrow$ N$_2$ (ice)  & ...  & ...  & ... & ... & 0.85 & ... & ... & ... \\
    CO (ice) $\rightarrow$ CO   & ...  & ... & ... & ... & ... & 855 & ... & ... \\
    N$_2$ (ice) $\rightarrow$ N$_2$  & ...  & ...  & ...  & ... & ... & 800 & ... & ... \\
    
    {\color[gray]{.4}CO (ice) + h$\nu \rightarrow$ CO}   & {\color[gray]{.4}...}  & {\color[gray]{.4}...}  & {\color[gray]{.4}...}  & {\color[gray]{.4}...}  & {\color[gray]{.4}...}  & {\color[gray]{.4}...}  & {\color[gray]{.4} $1.4\times10^{-3}$} & {\color[gray]{.4}...}  \\
    {\color[gray]{.4}N$_2$ (ice) + h$\nu \rightarrow$ N$_2$}   & {\color[gray]{.4}...}  & {\color[gray]{.4}...}  & {\color[gray]{.4}...}  & {\color[gray]{.4}...}  & {\color[gray]{.4}...}  & {\color[gray]{.4}...}  & {\color[gray]{.4} $2.1\times10^{-3}$} & {\color[gray]{.4}...}  \\
        
    CO + h$\nu$ $\rightarrow$ C + O	 & ... & ...  & ...  & ... & ... & ... & ... & $4.4\times10^{-7}$ \\
    N$_2$ + h$\nu$ $\rightarrow$ 2 N  & ... & ...  & ... & ...  & ... & ... & ... & $3.9\times10^{-7}$ \\ 
    \hline
\end{tabular}
\tablefoot{Equations for the reaction rate coefficients or reaction rates can be found in Appendix~\ref{ap:chemmodel}. Photodesorption processes are shown in grey and are only considered in model CH-PD. For photodissociation the unshielded rates are listed. \\ \tablefoottext{a} \mbox{Cosmic} ray ionization rate taken from \citet{Cravens1978}. \tablefoottext{b} \mbox{Values} taken from the \textsc{Rate}12 release of the UMIST database for Astrochemistry \citep{McElroy2013}. \tablefoottext{c} \mbox{Lower} limits for the sticking coefficients taken from \citet{Bisschop2006}. \tablefoottext{d} \mbox{Binding} energies for pure ices taken from \citet{Bisschop2006}. \tablefoottext{e} \mbox{Photodesorption} yields. For CO, the yield is taken from  \citet{Paardekooper2016} for CO ice at 20~K. For N$_2$, the result from \citet{Bertin2013} for mixed ices with CO:N$_2$ = 1:1 in protoplanetary disks is used. The yield for CO under these conditions is similar to the one reported by \citet{Paardekooper2016}. \tablefoottext{f} \mbox{Unattenuated} photodissociation rates for the adopted radiation field at a disk radius of 25 AU. Unshielded photodissociation rates for CO are taken from \citet{Visser2009} and for N$_2$ from \citet{Li2013} and \citet{Heays2014}.} 
\end{table*}

\begin{table}
\addtolength{\tabcolsep}{-2pt}
\caption{Overview of models and adopted parameters.
\label{tab:Models}} 
\centering
\begin{tabular}{l c c c c c c c c}
    \hline\hline
    \\[-.3cm]
    
    Model & $\chi$ \tablefootmark{a} & $E_\mathrm{b}$(CO) \tablefootmark{b} & $E_\mathrm{b}$(N$_2$) \tablefootmark{b} & $\zeta_{\mathrm{CR}}$ \tablefootmark{c}   & Photo- \\ 
    & & [K] & [K] & [s$^{-1}$] & desorption  \\
    \hline 
    \\[-.3cm]
    CH & 0.2 & 855 & 800 & $1.2 \times 10^{-17}$ &   \\
    CH-Eb1 & 0.2 & 1150 & 1000 & $1.2 \times 10^{-17}$ &   \\
    CH-Eb2 & 0.2 & 1150 & 800 & $1.2 \times 10^{-17}$ &   \\
    CH-CR1 & 0.2 & 855 & 800 & $1.0 \times 10^{-19}$ &   \\
    CH-CR2 & 0.2 & 855 & 800 & $5.0 \times 10^{-17}$ &   \\
    CH-PD & 0.2 & 855 & 800 & $1.2 \times 10^{-17}$ & yes \\
    CH-$\chi$0.8 & 0.8 & 855 & 800 & $1.2 \times 10^{-17}$ & \\
    \hline
\end{tabular}
\tablefoot{\tablefoottext{a} \mbox{Large} grain settling parameter. \tablefoottext{b} \mbox{Binding} energy. \tablefoottext{c} \mbox{Cosmic} ray ionization rate.}
\addtolength{\tabcolsep}{+2pt}
\end{table}
 
Prediction of the N$_2$H$^+$ abundance itself requires solving a chemical model. To avoid uncertainties associated with full chemical network models, a reduced chemical network, incorporating the key processes affecting the N$_2$H$^+$ abundance, including the freeze-out and thermal desorption of CO and N$_2$, is adopted. This network is similar to that used by \citet{Jorgensen2004a} for protostellar envelopes, but with freeze-out, thermal desorption and photodissociation of CO and N$_2$ included (see Fig. \ref{fig:ChemicalNetwork}). It resembles the analytical approach applied by \citet{Aikawa2015}. The most important aspects are described below and a more detailed description can be found in Appendix~\ref{ap:chemmodel}. 

Incorporation of CO and N$_2$ destruction by photodissociation in the surface and outer layers of the disk is necessary because depletion of the parent molecule, and a possible change in N$_2$/CO ratio, may affect the N$_2$H$^+$ abundance. For CO and N$_2$, photodissociation occurs through line absorption, and shielding by H$_2$ and self-shielding are important. For CO, photodissociation cross sections and shielding functions were taken from \citet{Visser2009}, and for N$_2$ from \citet{Li2013} and \citet{Heays2014}. For a given radiation field, both photodissociation rates are accurate to better than 20\%, and the difference in  unshielded rates ($2.6\times10^{-10}$ versus $1.7\times10^{-10}$ s$^{-1}$ in the general interstellar radiation field) turns out to be significant. Note that gas-phase formation of CO and N$_2$ are ignored, such that the model predicts a steep cutoff in the gas-phase abundances in the disk atmosphere. However, this should not affect the freeze-out and desorption balance around the snow surfaces, as they are located deeper within in the disk. 

The system of ordinary differential equations dictating the reduced chemistry, was solved using the python function \texttt{odeint}\footnote{The function \texttt{odeint} is part of the \texttt{SciPy} package (http://www.scipy.org/) and uses \texttt{lsoda} from the FORTRAN library \texttt{odepack}.} up to a typical disk lifetime of $10^{6}$ yr. As an initial condition, all CO and N$_2$ is considered to be frozen out, while all other abundances (except H$_2$) are set to zero. In Sect.~\ref{sec:Abundance} the effect of CO and N$_2$ abundances, and the N$_2$/CO ratio, is studied by varying the total, i.e. gas plus ice, abundances between $10^{-7}$ and $10^{-4}$ (with respect to H$_2$) such that the N$_2$/CO ratio ranges between 0.01 and 100. We will refer to these models as model ``CH'' (simple CHemical network). The adopted parameters are listed in Table~\ref{tab:RateCoefficients}.

\begin{figure*}
\centering
\includegraphics[width=17cm,trim={0 16.2cm 0cm 1.1cm},clip]{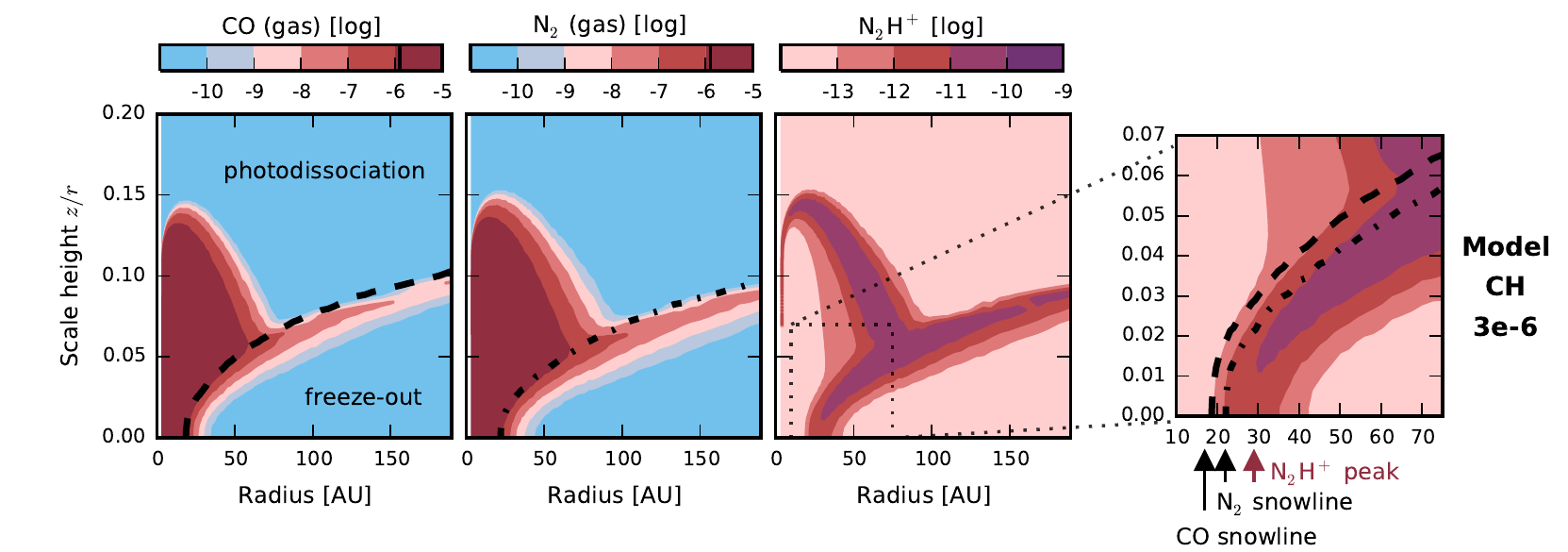}
\caption{Distributions of CO gas, N$_2$ gas and N$_2$H$^+$ in the simple chemical model (model \mbox{CH}) with CO and N$_2$ abundances of $3 \times 10^{-6}$. To focus on the region around the CO snow surface, the vertical scale is limited to a scale height $z/r \leq$ 0.2. The rightmost panel highlights the region where N$_2$H$^+$ is present near the disk midplane. The dashed and dash-dotted contours represent the CO and N$_2$ snow surfaces, respectively, and the corresponding midplane snowlines are indicated by arrows below the horizontal axis of the rightmost panel. The midplane radius with the highest N$_2$H$^+$ abundance is marked with a red arrow.}
\label{fig:N2H+distribution}
\end{figure*}

The temperature at which a molecule freezes out depends on the gas density and on the binding energy for each molecule, $E_\mathrm{b}$. In the fiducial FD and CH models binding energies for pure ices are used. When in contact with water ice, the CO and N$_2$ binding energies are higher. Recent results from \citet{Fayolle2016} show that, as long as the ice morphology and composition are equivalent for both CO and N$_2$,  the ratio of the binding energies remains the same ($\sim$0.9). The effect of different binding energies will be studied in Sect.~\ref{sec:Eb} by adopting values of 1150~K and 1000~K (model \mbox{CH-Eb1}) and 1150~K and 800~K (model \mbox{CH-Eb2}), for CO and N$_2$, respectively. The former values are for both CO and N$_2$ on a water ice surface \citep{Garrod2006}, i.e. representing a scenario in which all ices evaporate during disk formation and then recondense. The latter model represents a situation in which CO is in contact with water ice, while N$_2$ resides in a pure ice layer. 

Another important parameter in the simple chemical model is the cosmic ray ionization rate, since it controls the H$_3^+$ abundance, important for formation of N$_2$H$^+$. Based on modeling of HCO$^+$ and N$_2$H$^+$ line fluxes and spatially resolved emission, \citet{Cleeves2015} have suggested that the cosmic ray ionization rate in TW Hya is very low, of order \mbox{$10^{-19}$ s$^{-1}$.} The importance of the cosmic ray ionization rate is addressed in Sect.~\ref{sec:ZetaCR} by adopting values of \mbox{$\zeta = 1 \times 10^{-19}$ s$^{-1}$} (CH-CR1) and \mbox{$\zeta = 5 \times 10^{-17}$ s$^{-1}$} (CH-CR2), as also used by \citet{Aikawa2015} in their study of N$_2$H$^+$.

An overview of all CH models is given in Table~\ref{tab:Models}.


\subsection{Line radiative transfer}

Emission from the N$_2$H$^+$ $J$ = 4--3 (372~GHz), $J$ = 3--2 (279 GHz) and $J$ = 1--0 (93 GHz) transitions were simulated with the radiative transfer code LIME \citep[LIne Modeling Engine,][]{Brinch2010} assuming a distance, inclination and position angle appropriate for TW Hya; 54 pc, 6$\degr$ and 155$\degr$, respectively \citep{Hughes2011,Andrews2012}. These are the same values as adopted by Q13. The LIME grid was constructed such that the grid points lie within and just outside the region where the N$_2$H$^+$ abundance $> 1\times10^{-13}$. In the disk region where N$_2$H$^+$ is predicted, the gas density is larger than the $J$=4--3 critical density of $\sim8\times10^6$~cm$^{-3}$ (see Fig.~\ref{fig:PhysicalStructure}), so to reduce CPU time, models were run in LTE. The simulated images were convolved with a 0\farcs63 $\times$ 0\farcs59 beam, similar to the reconstructed beam of Q13, and a 0\farcs2 $\times$ 0\farcs2 beam to anticipate future higher spatial resolution observations.  For the \mbox{$J$ = 4--3} transition, the line profiles and the integrated line intensity profiles were compared to the observational data reduced by Q13.


\section{Results} \label{sec:Results}


\subsection{N$_2$H$^+$ distribution and emission} \label{sec:CompareModels}

Figure \ref{fig:N2H+distribution} shows the distribution of CO gas, N$_2$ gas and N$_2$H$^+$ as predicted by the simple chemical model (model CH). Abundance refers to fractional abundance with respect to H$_2$ throughout this work. CO and N$_2$ are frozen out in the disk midplane and destroyed by UV photons higher up in the disk. The snow surface is defined as the position in the disk where the gas-phase and ice abundances become equal (dashed and dash-dotted contours in Fig.~\ref{fig:N2H+distribution}, left panels), and the snowline is the radius at which this happens in the midplane. For the physical structure and fiducial binding energies adopted, the CO snowline is then located at 19~AU which corresponds to a temperature for both the gas and dust of \mbox{$\sim$20~K}. This is smaller than the snowline location of 30~AU (corresponding to 17~K) as inferred by Q13, but in good agreement with recent results from \citet{Zhang2016} who directly detect the CO snowline around 17~AU using \element[][13]C\element[][18]O observations.

Although the N$_2$H$^+$ abundance starts to increase at the midplane CO snowline, it peaks \mbox{$\sim$10~AU} further out (red arrow in Fig.~\ref{fig:N2H+distribution}, rightmost panel). It thus seems that the reduction in CO gas abundance at the snowline is not sufficient to allow N$_2$H$^+$ to be abundant, but that an even higher level of depletion is required to favor N$_2$H$^+$ formation over destruction. On the other hand, very low fractions of N$_2$ in the gas phase are sufficient to allow N$_2$H$^+$ formation, extending the N$_2$H$^+$ layer beyond the N$_2$ snow surface. In addition to the expected N$_2$H$^+$ layer, N$_2$H$^+$ is predicted to be abundant in a layer higher up in the disk where the N$_2$ abundance in the gas phase exceeds that of CO due to a slightly lower photodissociation rate of N$_2$ as compared with CO. The presence of N$_2$H$^+$ in the surface layers is also seen in full chemical models \citep{Walsh2010,Cleeves2014,Aikawa2015} and its importance is further discussed in Sect.~\ref{sec:SurfaceLayer}.

\begin{figure}
\centering
\includegraphics[trim={0 6.8cm 0cm 1cm},clip]{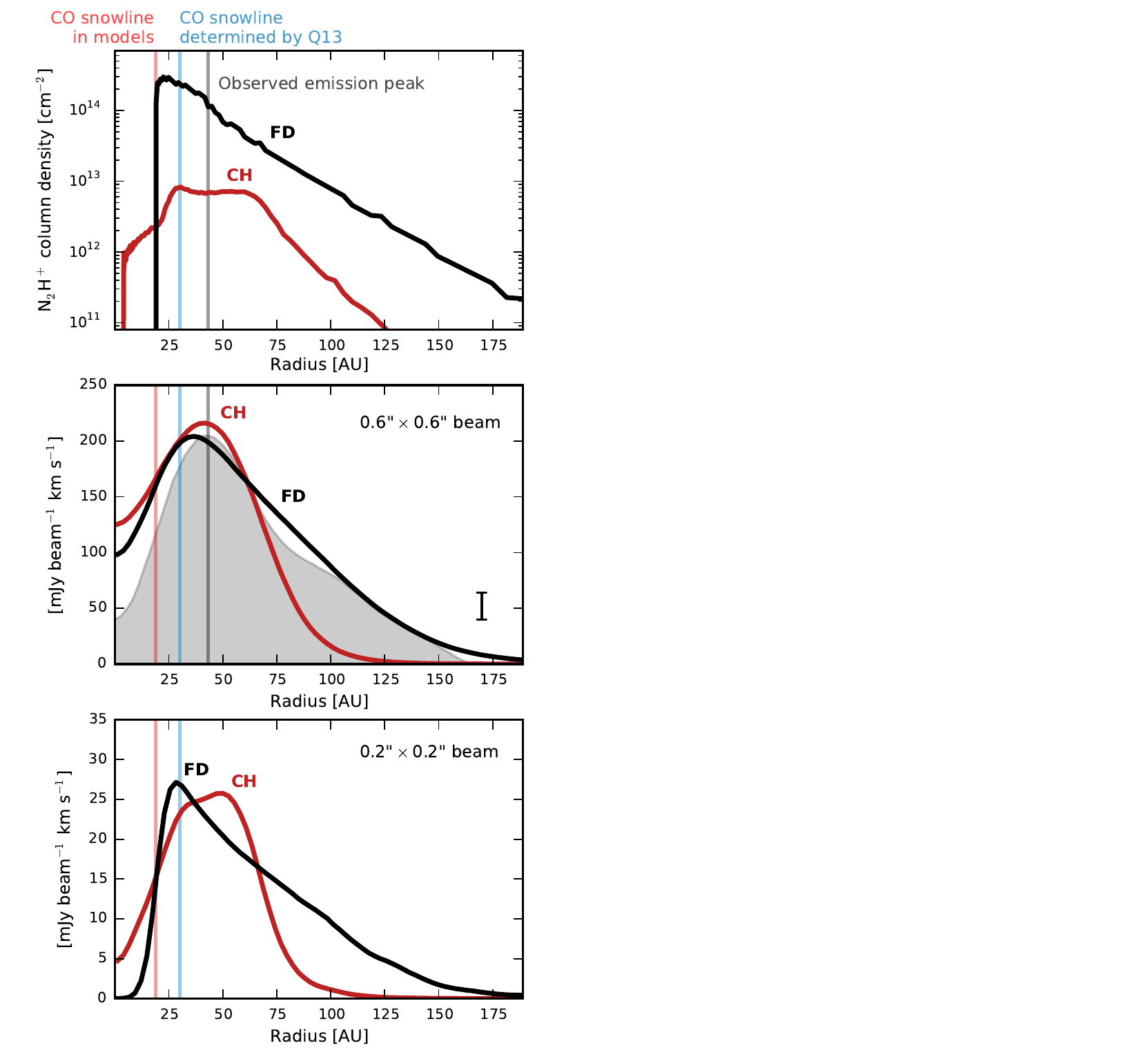}
\caption{N$_2$H$^+$ column density profile (top panel) and simulated \mbox{$J$ = 4--3} line emission (middle and bottom panel) for the N$_2$H$^+$ distributions predicted by the simple chemical model with CO and N$_2$ abundances of $3\times10^{-6}$ (model CH; red lines) and a model incorporating only freeze-out and desorption (model FD; black lines). Integrated line intensity profiles are shown after convolution with a $0\farcs63\times0\farcs59$ beam (middle panel) or a $0\farcs2\times0\farcs2$ beam (bottom panel). Observations by Q13 are shown in grey in the middle panel with the 3$\sigma$-error depicted in the lower right corner. The vertical grey line marks the position of the observed emission peak. The vertical blue line indicates the position of the midplane CO snowline inferred from these observations by Q13, while the red line indicates the location of the midplane CO snowline in the models.} 
\label{fig:Emission}
\end{figure}

The results from the simple chemical model thus deviate from the expectation that N$_2$H$^+$ is most abundant in a layer directly outside the CO snowline, as can also be seen from the radial column density profiles in Fig.~\ref{fig:Emission} (top panel). When considering only freeze-out and desorption (model FD) and assuming a constant N$_2$H$^+$ abundance of $3\times10^{-10}$ between the CO snow surface and the 0.05\% contour for N$_2$ gas, the N$_2$H$^+$ column density peaks only 2 AU outside the snowline. On the contrary, in model CH this peak is located 11 AU further out in the disk, at the snowline location derived by Q13. In addition, the column density profile for model CH is flatter due to the N$_2$H$^+$ surface layer. 

\begin{figure}
\sidecaption
\includegraphics{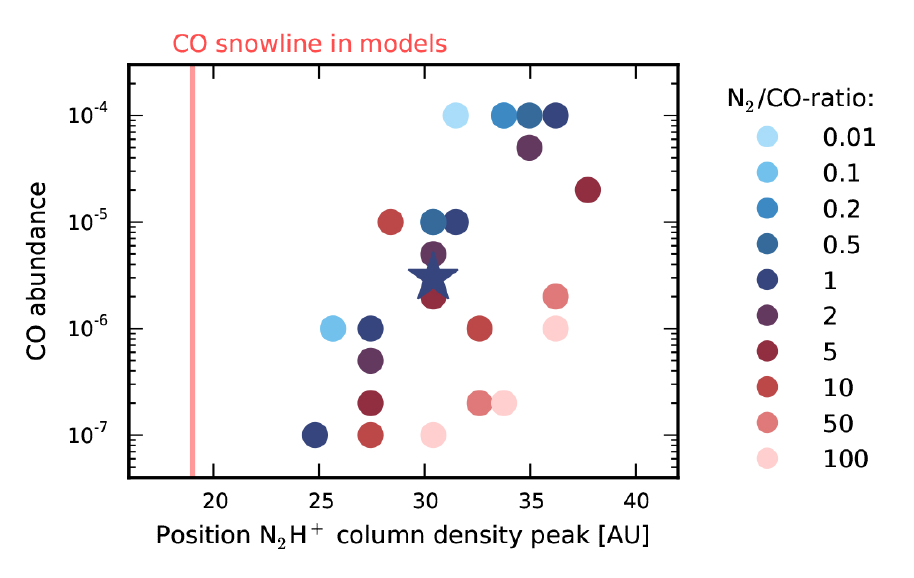}
\caption{Position of the N$_2$H$^+$ column density peak in model CH for different CO and N$_2$ abundances. The best-fit model with abundances of $3\times10^{-6}$, as shown in Fig.~\ref{fig:N2H+distribution}, is indicated by a star and the color of the symbols represents the value of the N$_2$/CO ratio. The vertical red line marks the location of the CO snowline in the models.}
\label{fig:AbundanceColDens}
\end{figure}

In order to determine whether this difference in N$_2$H$^+$ distribution is large enough to cause different emission profiles, emission from the N$_2$H$^+$ $J$~=~4--3 (372~GHz) transition was simulated. Model FD fits the observed emission peak reasonably well for an N$_2$H$^+$ abundance of $3 \times 10^{-10}$, although the simulated emission peak is located 7~AU closer to the star than observed. Variations in the assumed N$_2$H$^+$ abundance only affect the intensity, but not the position of the peak. On the other hand, model CH can reproduce the position of the emission peak for a CO and N$_2$ abundance of $3 \times 10^{-6}$ (Fig.~\ref{fig:Emission}, middle panel). The underprediction of the emission in the outer disk is further discussed in Sect.~\ref{sec:OuterEdge}. The difference between the models becomes more prominent at higher spatial resolution (Fig.~\ref{fig:Emission}, bottom panel). In that case, model FD predicts the emission peak 10~AU outside the snowline (instead of 17~AU), while this is 30~AU for model CH (instead of 24~AU) due to the flattened column density profile. An N$_2$H$^+$ column density peaking at 30~AU, 11~AU outside the snowline, can thus reproduce the observed emission peak, which is in agreement with Q13, unlike a column density profile peaking directly at the CO snowline. However, this is only the case for a low CO and N$_2$ abundance of $3\times10^{-6}$, as discussed further below.


\subsection{Influence of CO and N$_2$ abundances}\label{sec:Abundance}

To examine whether the exact amount of CO present in the gas phase is more important for the N$_2$H$^+$ distribution than the location of the CO snowline, as suggested above, the total CO and N$_2$ abundances in the simple chemical network were varied. Changing the CO abundance does not influence the N$_2$H$^+$ distribution via temperature changes since the gas and dust are coupled in the region where N$_2$H$^+$ is present (see Sect.\ref{sec:Physicalmodel} and Fig.~\ref{fig:PhysicalStructure}). Furthermore, recall that the location of the midplane CO snowline does not depend on abundance and thus remains at 19~AU for all models which adopt the fiducial binding energy. The position of the N$_2$H$^+$ column density peak, however, turns out to move further away from the snowline with increasing CO abundance (Fig.~\ref{fig:AbundanceColDens}). This reinforces the idea that the gas-phase CO abundance remains too high for N$_2$H$^+$ to be abundant after the 50\% depletion at the snowline. Instead, N$_2$H$^+$ peaks once the amount of CO in the gas phase drops below a certain threshold, which is reached further away from the snowline for higher CO abundances. This is in agreement with the conclusions from \citet{Aikawa2015}. 

Moreover, the position of the column density peak depends also on the N$_2$ abundance. For a fixed CO abundance, the position of the maximum N$_2$H$^+$ column density shifts outward with increasing N$_2$ abundance, since the amount of gas-phase N$_2$ remains high enough for efficient N$_2$H$^+$ formation at larger radii. The N$_2$H$^+$ distribution thus strongly depends on the amount of both CO and N$_2$ present in the gas phase, with the column density peaking \mbox{6--18 AU} outside the CO snowline for different abundances.


\subsection{Importance of the N$_2$H$^+$ surface layer}\label{sec:SurfaceLayer}

\begin{figure}
\sidecaption
\includegraphics[width=0.5\textwidth]{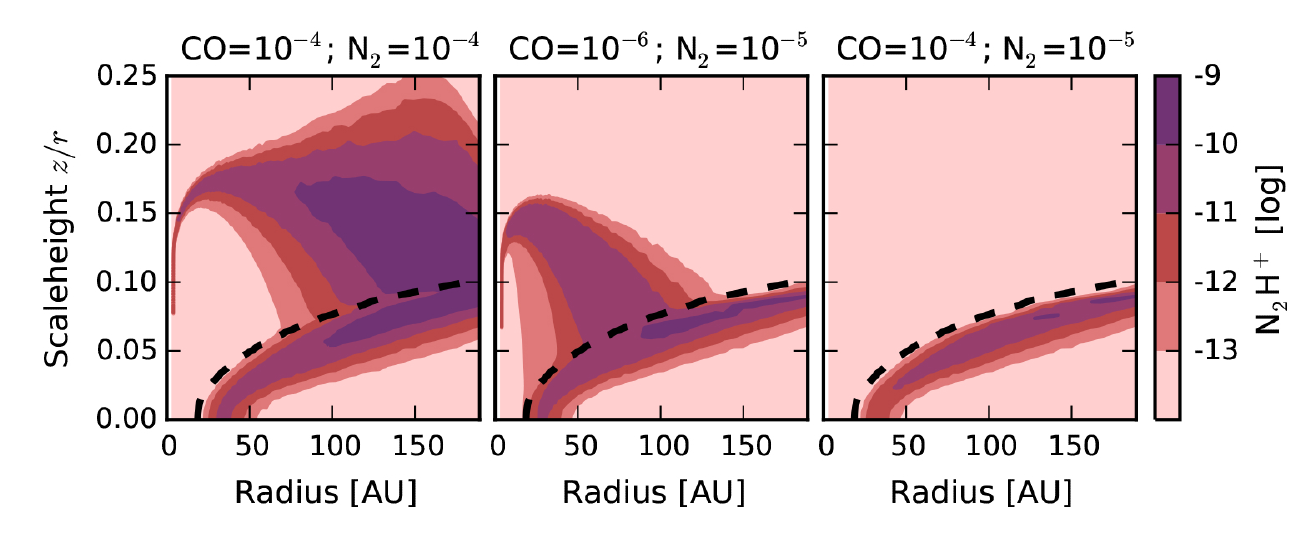}
\caption{Distribution of N$_2$H$^+$ in the simple chemical model (model CH) for different N$_2$ and CO abundances as listed above the panels. To focus on the region around the CO snow surface, the vertical scale is limited to a scale height $z/r \leq$ 0.25. The dashed contour represents the CO snow surface.} 
\label{fig:N2H+_abundances}
\end{figure}

Besides the expected N$_2$H$^+$ layer outside the CO snow surface, model CH also predicts a layer higher up in the disk where N$_2$H$^+$ is abundant as a result of a slightly lower N$_2$ photodissociation rate compared with CO. Since both molecules can self-shield, the photodissociation rates depend on molecular abundances. Therefore, the CO and N$_2$ abundances influence the shape of the N$_2$H$^+$ surface layer as shown in Fig.~\ref{fig:N2H+_abundances}. When N$_2$ is equally or more abundant than CO, N$_2$H$^+$ can survive in the region where CO is photodissociated but N$_2$ is still present. The higher the abundances, the closer to the disk surface a sufficiently high column density is reached for efficient self-shielding and the more extended is the N$_2$H$^+$ surface layer (Fig.~\ref{fig:N2H+_abundances}, left panel). The inner boundary of the surface layer is set where CO photodissociation ceases to be effective. For lower CO and N$_2$ abundances, photodissociative photons can penetrate deeper into the disk, and the N$_2$H$^+$ surface layer is located closer to the star (Fig.~\ref{fig:N2H+_abundances}, middle panel). The layer does not extend to the disk outer radius any longer because most N$_2$ is now photodissociated in the outer regions. Finally, when CO is more abundant than N$_2$, the surface layer decreases, until for N$_2$/CO $\lesssim$ 0.2 CO becomes abundant enough everywhere above the snow surface to shift the balance towards N$_2$H$^+$ destruction (Fig.~\ref{fig:N2H+_abundances}, right panel). 

\begin{figure*}
\vspace{-0.4cm}
\centering
\includegraphics[width=17cm,trim={0 12.3cm 0cm 1.3cm},clip]{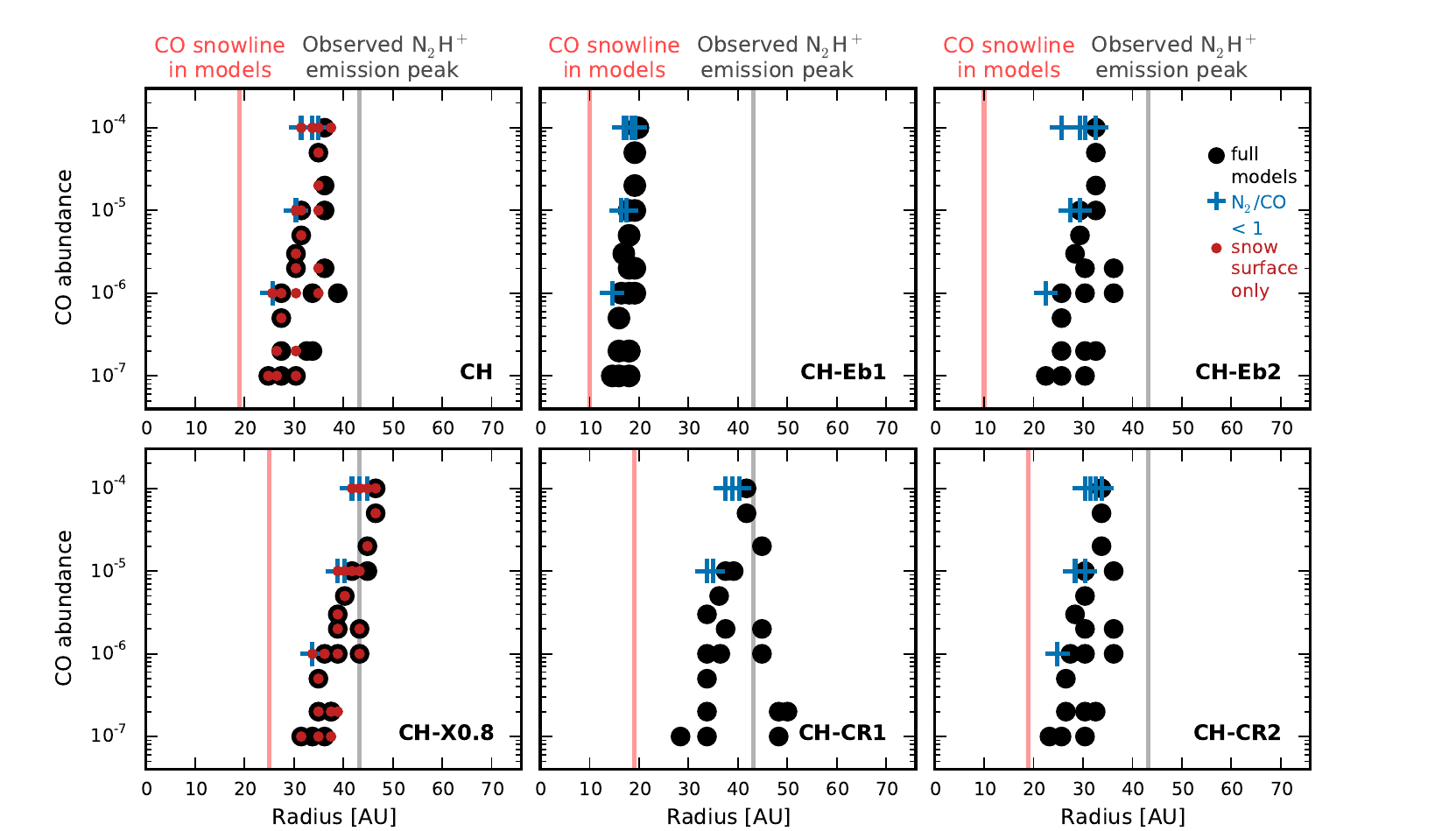}
\caption{Position of the N$_2$H$^+$ column density peak in the different models (listed in the lower right corner of each panel) for different CO and N$_2$ abundances. From left to right and top to bottom: the fiducial models (CH), models with both CO and N$_2$ binding energies increased (CH-Eb1), models with only CO binding energy increased (CH-Eb2), models with large grains settled to only 80\% of small grain scale height (CH-$\chi$0.8), models with a lower cosmic ray ionization rate ($1\times10^{-19}$~s$^{-1}$; CH-CR1) and models with a higher cosmic ray ionization rate ($5\times10^{-17}$~s$^{-1}$; CH-CR2). Models with N$_2$/CO ratios $<$ 1 are highlighted with blue plus signs. Red circles in the left panels represent the snow surface only models, i.e. N$_2$H$^+$ removed above the CO snow surface. The red lines mark the location of the CO snowline in the models. The grey line indicates the position of the observed emission peak.} 
\label{fig:ColumndensityPeak}

\includegraphics[width=17cm,trim={0 12.3cm 0cm 1.3cm},clip]{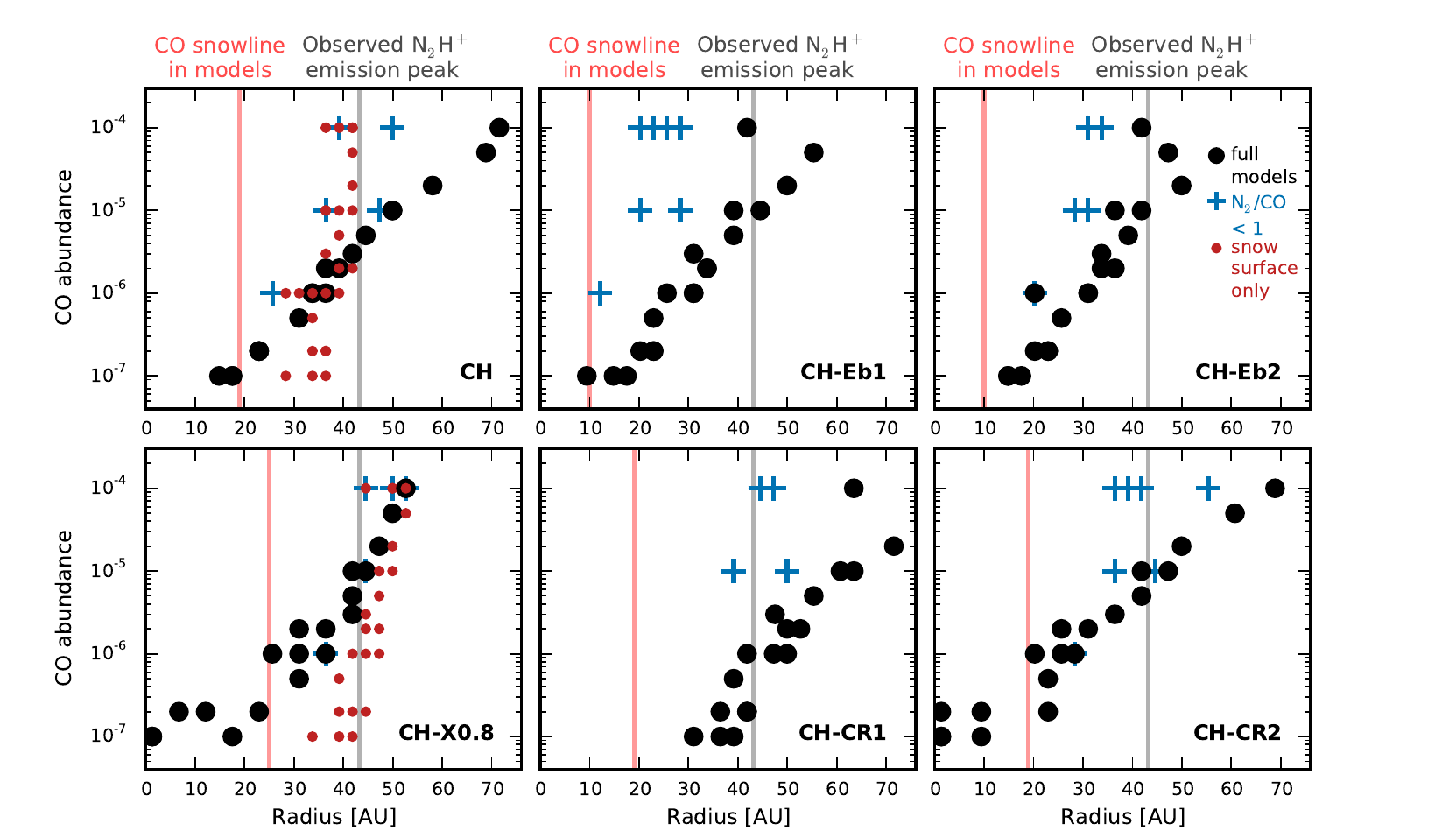}
\caption{As Fig.~\ref{fig:ColumndensityPeak}, but for the position of the simulated N$_2$H$^+$ $J$=4--3 emission peak after convolution with a $0\farcs63\times0\farcs59$ beam. } 
\label{fig:EmissionPeak}
\vspace{-0.1cm}
\end{figure*}

To address the influence of the N$_2$H$^+$ surface layer, \mbox{$J$=4--3} lines were simulated for model CH with different CO and N$_2$ abundances with the CO snow surface set as an upper boundary. In other words, no N$_2$H$^+$ is present above the CO snow surface in these ``snow surface only'' models. Removing the N$_2$H$^+$ surface layer hardly  affects the position of the column density peak (Fig.~\ref{fig:ColumndensityPeak}, top left panel), suggesting that the offset between N$_2$H$^+$ and CO snowline is not caused by the surface layer but rather is a robust chemical effect. The emission, however, is strongly influenced by the surface layer (Fig.~\ref{fig:EmissionPeak}, top left panel). In the full CH models, the emission peak is shifted away from the snowline for higher CO abundances by up to $\sim$~50 AU, while in the snow surface only models, the emission traces the column density peak with an offset related to the beam size. Only for CO abundances~$\sim10^{-6}$ or N$_2$/CO ratios $\lesssim$ 1 (blue plus signs in Fig.~\ref{fig:EmissionPeak}), does the emission trace the column density in the full models, and only for even lower CO abundances ($\sim10^{-7}$) does the emission peak at the snowline. In addition to the N$_2$H$^+$ column density offset, the relation between CO snowline and N$_2$H$^+$ emission is thus weakened even more in models with N$_2$/CO $\gtrsim$ 0.2 due to the presence of an N$_2$H$^+$ surface layer that causes the emission to shift outward. 

Furthermore, the N$_2$H$^+$ surface layer contributes significantly to the peak integrated intensity. This intensity shows a linear correlation with the N$_2$/CO ratio, but the difference of \mbox{$\sim$600 mJy beam$^{-1}$ km s$^{-1}$} (for the $0\farcs63\times0\farcs59$ beam) between models with a N$_2$/CO ratio of 0.01 and 100 reduces to only \mbox{$\sim$100 mJy beam$^{-1}$ km s$^{-1}$} in the snow surface only models (see Fig.~\ref{fig:EmissionIntensity}). For the TW Hya physical model adopted, a surface layer of N$_2$H$^+$, in addition to the midplane layer outside the CO snow surface, seems necessary to reproduce the observed integrated peak intensity. This is in agreement with \citet{Nomura2016}, who suggest that the N$_2$H$^+$ emission in TW Hya originates in the disk surface layer based on the brightness temperature.

\subsection{Influence of CO and N$_2$ binding energies}\label{sec:Eb}

The location of the CO snowline depends on the CO binding energy. To address whether the offset between N$_2$H$^+$ and CO snowline is a result of the adopted binding energies, models were run with a higher CO binding energy (1150 K), i.e. assuming CO on a water ice surface (model CH-Eb2). As the amount of N$_2$ also influences the N$_2$H$^+$ distribution, models were run with a higher binding energy for both CO and N$_2$ (1150 and 1100 K, respectively) as well (model CH-Eb1). The position of the N$_2$H$^+$ column density and emission peak for different CO and N$_2$ abundances are shown in the top middle and top right panels of Figs.~\ref{fig:ColumndensityPeak} and \ref{fig:EmissionPeak}, respectively. When the binding energy is increased for both species (model CH-Eb1), the results are similar to before. The N$_2$H$^+$ column density peaks \mbox{5--9 AU} outside the CO snowline, and the emission peak shifts to even larger radii with increasing CO abundance when an N$_2$H$^+$ surface layer is present (black circles in Fig.~\ref{fig:EmissionPeak}). Increasing only the CO binding energy, i.e. shifting the CO snowline inward but not affecting the N$_2$ snowline (model CH-Eb2), results in the N$_2$H$^+$ column density to peak \mbox{12--26 AU} from the CO snowline. The emission peaks, however, stay roughly at the same radii for both models, thus better tracing the column density maximum when the CO and N$_2$ snowlines are further apart. The peak integrated intensities are similar for all three sets of binding energies.  

The N$_2$H$^+$ column density thus peaks outside the CO snowline for all binding energies tested, and the offset is largest when the CO and N$_2$ snowline are furthest apart. The offset between snowline and emission peak is roughly independent of the binding energies, except for CO abundances of $\sim10^{-4}$. Therefore, a degeneracy exists between the peak position of the emission and the column density.


\subsection{Influence of the cosmic ray ionization rate}\label{sec:ZetaCR}

The cosmic ray ionization rate controls the H$_3^+$ abundance, and may therefore have an effect on the N$_2$H$^+$ distribution. To address the importance of the cosmic ray ionization rate, model CH was run with \mbox{$\zeta = 5 \times 10^{-17}$ s$^{-1}$} (CH-CR2), as also used by \citet{Aikawa2015} in their study of N$_2$H$^+$, and \mbox{$\zeta = 1 \times 10^{-19}$ s$^{-1}$} (CH-CR1), as suggested by \citet{Cleeves2015}. The results for the N$_2$H$^+$ column density and \mbox{$J$=4--3} emission are presented in Figs.~\ref{fig:ColumndensityPeak} and \ref{fig:EmissionPeak}, respectively (bottom middle and right panels). The trends seen for the position of the column density and emission peak are roughly the same as for the fiducial models with \mbox{$\zeta = 1.2 \times 10^{-17}$ s$^{-1}$}, although both offsets are $\sim$10~AU larger for the lowest cosmic ray ionization rate. The very small radius at which the emission peaks for model CH-CR2 with a CO abundance of $\sim10^{-7}$ is due to a combination of a higher N$_2$H$^+$ abundance in the inner few tens of AU as compared to models with higher CO abundance and a $0\farcs6$ ($\sim$32~AU) beam. The strongest effect of the cosmic ray ionization rate is on the strength of the peak integrated intensity. Models CH-CR2 predict a higher peak integrated intensity than observed, while N$_2$ needs to be more than two orders of magnitude more abundant than CO to be consistent with the low cosmic ray ionization rate of \mbox{$10^{-19}$ s$^{-1}$} in models CH-CR1 (see Fig.~\ref{fig:Intensity_ZetaCR}). 

The cosmic ray ionization rate thus influences the distribution of N$_2$H$^+$ with respect to the snowline, with the column density peaking closest to the snowline for the highest values of $\zeta$ and the lowest CO abundances. However, the smallest offset remains 4~AU. 

\begin{figure}
\centering
\includegraphics[scale=0.9]{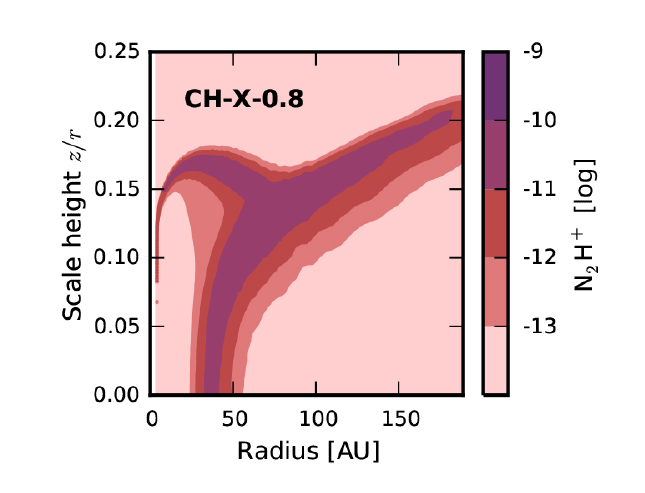}
\vspace{-0.2cm}
\caption{N$_2$H$^+$ distribution predicted by the simple chemical model for a physical structure with the large grains settled to only 80\% of the small grain scale height (model CH-$\chi$0.8). Abundances of $3\times10^{-6}$ are adopted for both CO and N$_2$.} 
\label{fig:N2H+Grainsettling}
\end{figure}


\subsection{Influence of grain settling}\label{sec:GrainSettling}

In the physical model adopted so far, the large grains have settled toward the disk midplane. The distribution of the dust is important because it affects the UV penetration and the disk thermal structure, which is determined by the processing of UV radiation by the dust particles. Since the location of the CO snow surface is temperature dependent, grain settling may indirectly influence the location of the CO snowline. To examine whether this also influences the relation between N$_2$H$^+$ and the snowline, a physical model in which the large grains have only settled to 80\% of the small grain scale height is used. The N$_2$H$^+$ distribution predicted by the simple chemical model for CO and N$_2$ abundances of $3\times10^{-6}$ is presented in Fig.~\ref{fig:N2H+Grainsettling}. The CO snow surface is now located higher up in the disk as a consequence of the shallower temperature gradient near the midplane. In other words, the temperature stays below the CO freeze-out temperature at larger scale heights. The resulting increase in the N$_2$H$^+$ column just outside the snowline in combination with the smaller N$_2$H$^+$ surface layer, reduces the contribution of this layer. This is for instance reflected in the peak integrated intensity; the difference between full models and snow surface only models is now only a factor of $\sim$2 instead \mbox{of $\sim$5}. 

Figures~\ref{fig:ColumndensityPeak} and \ref{fig:EmissionPeak} (bottom left panels) show what this means for the positions of the N$_2$H$^+$ column density and emission peaks. Due to the different temperature structure, the CO snowline is located at 25~AU, but the N$_2$H$^+$ column density still peaks 6--22~AU further out. However, the offset between column density and emission peak is now different. The emission does trace the column density for CO abundances higher than $\sim5\times10^{-6}$, while for lower abundances the emission peaks at smaller radii than the column density. Again, when the surface layer is removed, the emission roughly traces the column density for all CO and N$_2$ abundances. 

Thus, the N$_2$H$^+$ emission seems not only sensitive to the chemical conditions, but also the physical conditions in the disk and the UV penetration. Depending on the degree of grain settling the emission traces the column density for different CO abundances, although the N$_2$H$^+$ column density peaks outside the CO snowline in all models. 

\begin{figure}
\centering
\includegraphics[width=17cm,trim={0 13.6cm 0cm 1.5cm},clip]{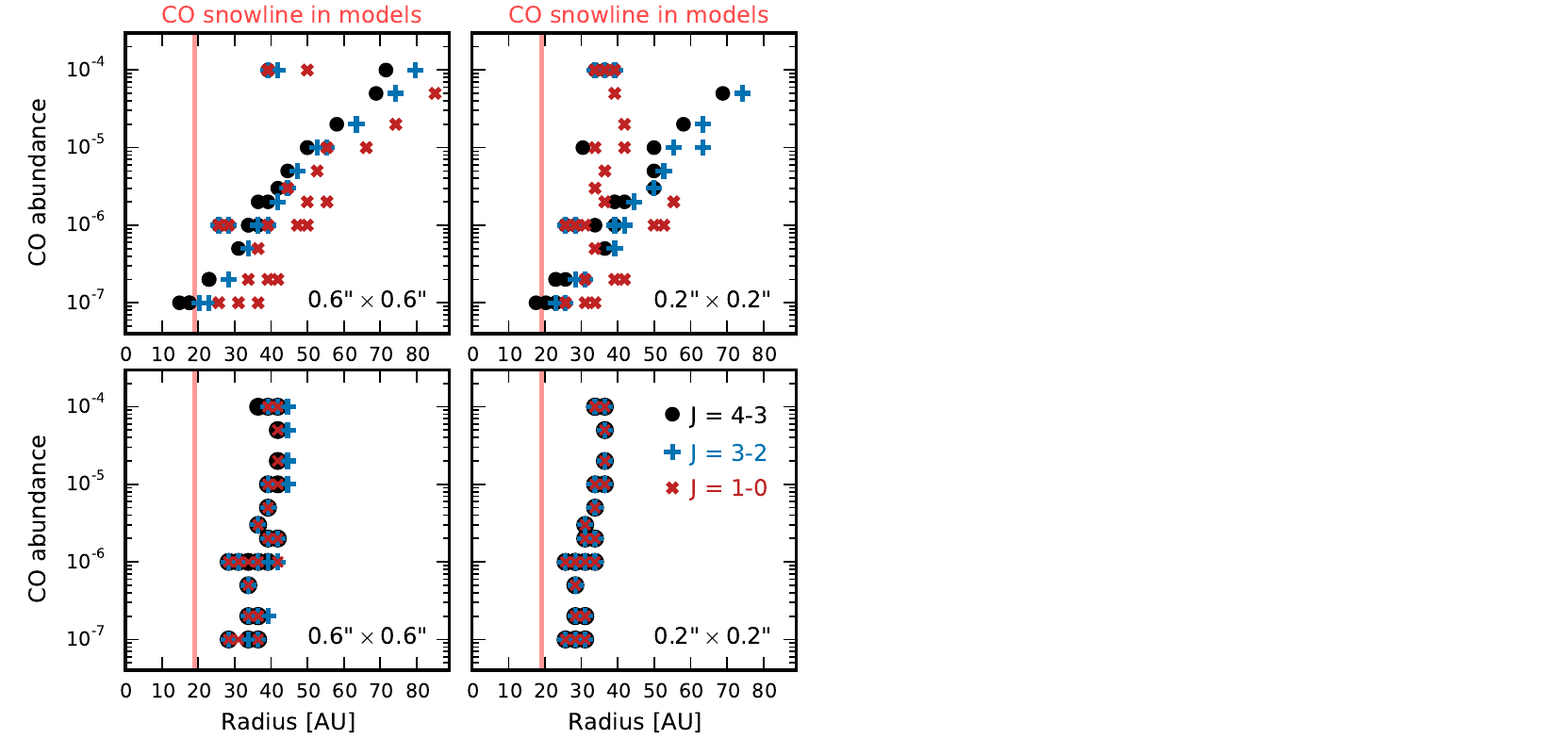}
\caption{Position of the N$_2$H$^+$ $J$=4--3 (black circles), $J$=3--2 (blue plus signs) and $J$=1--0 (red crosses) emission peaks for different CO and N$_2$ abundances in the simple chemical model (model CH) (top panels) and the corresponding snow surface only models, i.e. N$_2$H$^+$ removed above the CO snow surface (bottom panels). The emission is convolved with a $0\farcs63\times0\farcs59$ beam (left panels) or  $0\farcs2\times0\farcs2$ beam (right panels). The red lines mark the location of the CO snowline in the models.} 
\label{fig:Transitions}
\end{figure}


\subsection{Constraints provided by multiple N$_2$H$^+$ transitions}\label{sec:Transitions}

For N$_2$/CO ratios larger than $\sim$0.2, the simple chemical network predicts that N$_2$H$^+$ is also abundant in a surface layer above the CO snow surface. The presence of this surface layer significantly influences the N$_2$H$^+$ \mbox{$J$=4--3} emission and complicates the relationship between N$_2$H$^+$ and the CO snowline. To assess whether a different N$_2$H$^+$ transition would be better suited to trace the CO snowline, emission was simulated for the \mbox{$J$=3--2} (279~GHz) and \mbox{$J$=1--0} (93~GHz) transitions for models CH and CH-$\chi$0.8. The results for the position of the N$_2$H$^+$ emission peaks in model CH are shown in Fig.~\ref{fig:Transitions}. For the full models with N$_2$/CO~$>$~0.2, the emission peak shifts outward with decreasing transition frequency (Fig.~\ref{fig:Transitions}, top panels), while all transitions peak at a similar radius for the models where the N$_2$H$^+$ surface layer has been removed (Fig.~\ref{fig:Transitions}, bottom panels) or is not present. When the emission is convolved with a $0\farcs2\times0\farcs2$ beam, the \mbox{$J$=1--0} transition peaks in some models at smaller radii than the other transitions. That is because in these cases the structure caused by the surface layer can be resolved, revealing two components that are smeared into one broad feature by the $0\farcs63\times0\farcs59$ beam. Similar results are obtained for model CH-$\chi$0.8 (not shown). Observing multiple transitions thus seems to provide a good indication whether or not a surface layer of N$_2$H$^+$ is contributes to the emission, and thus how well the emission traces the column density. 

Although comparison of the emission-peak positions for different transitions may indicate the contribution of an N$_2$H$^+$ surface layer, no information is provided on how far the emission peak is then offset from the column density peak or actual CO snowline. To examine whether N$_2$H$^+$ line ratios may contribute to addressing this problem, the \mbox{$J$=4--3/$J$=3--2} and \mbox{$J$=4--3/$J$=1--0} ratios are calculated. Results for model CH and model CH-$\chi$0.8 with three different CO and N$_2$ abundances (as shown in Fig.~\ref{fig:N2H+_abundances}) are presented in Fig.~\ref{fig:Lineratios}. When the N$_2$H$^+$ surface layer is removed or not present at all, both line ratios are nearly constant throughout the disk at \mbox{$J$=4--3/$J$=3--2 $\approx$ 1.2} and \mbox{$J$=4--3/$J$=1--0 $\approx$ 20}. Only at $0\farcs2$ resolution does the \mbox{$J$=4--3/$J$=1--0} ratio increase in the inner $\sim$30~AU. In the fudicial model with the large grains settled to 20\% of the small grain scale height, both line ratios can distinguish between differently shaped N$_2$H$^+$ surface layers. The line ratios become more steep when the surface layer extends to about half the disk radius and increase in value for a surface layer extending to the disk outer radius. In model CH-$\chi$0.8, the surface layer contributes less to the emission and although the line ratios show an increase at around 40~AU when the surface layer is present, distinguishing differently shaped surface layers is not possible. N$_2$H$^+$ line ratios are thus sensitive to the distribution of N$_2$H$^+$, and together with the position of the different emission peaks, can provide modeling constraints and aid in constraining the location of the CO snowline. 

\begin{table}
\caption{Offset between the CO snowline and the N$_2$H$^+$ column density and $J$=4--3 emission peak in the different models.
\label{tab:Results}} 
\centering
\begin{tabular}{l c c c c c c c c}
    \hline\hline
    \\[-.3cm]
    
     & Offset & \multicolumn{4}{c}{\,\,\,Offset $J$=4--3 emission} \\ 
  Model & column density & $0\farcs63\times0\farcs59$ & $0\farcs2\times0\farcs2$ &  \\
   & AU & AU & AU \\
    \hline 
    \\[-.3cm]
    CH 		& \,\,\,6--18		& \,\,\,\,\,\,\,4--53 *		& \,\,\,\,\,\,\,2--50 *	\\
    CH-Eb1 & \,\,\,5--10	 	& \,\,\,\,\,\,\,2--45 *		& \,\,8--43 \\
    CH-Eb2 & 13--26			& \,\,5--40	                  & \,\,8--35 \\
    CH-CR1	& 10--31			& 12--53 		             & 12--55\\
    CH-CR2	& \,\,\,4--17		& \,\,\,\,\,\,\,2--50 *       & \,\,\,\,\,\,\,2--53 * \\
    CH-$\chi$0.8 & \,\,6--2		& \,\,\,\,\,\,\,1--28 *		& \,\,\,\,\,\,\,4--22 * \\
    \hline
\end{tabular}
\tablefoot{The CO snowline is located at 19 AU in models CH, CH-CR1 and CH-CR2, at 10 AU in models CH-Eb1 and CH-Eb2, and at 25 AU in models CH-$\chi$0.8. A value of ``0'' means coincidence with the CO snowline in the respective model. A star (*) indicates models for which the emission peaks inside the snowline for CO abundances $\leq2\times10^{-7}$.}
\end{table}


\subsection{Outer edge of N$_2$H$^+$ emission}\label{sec:OuterEdge}

So far, we have focused on the peak of the N$_2$H$^+$ emission and its relation to the CO snowline. The simple chemical model (model CH) produces a good fit to the emission peak, but underestimates the emission coming from the outer disk (further out than $\sim$60~AU). In this region, the density may have become low enough for UV radiation to penetrate the midplane and photodesorption to become effective. To address whether this can account for the observed emission, photodesorption is included in model CH-PD (see Fig.~\ref{fig:Emission_Photodesorption}). Although N$_2$H$^+$ is now present in the midplane at radii larger than $\sim$~60~AU and this results in an increase in the column density at these radii, the $\sim$10~mJy~beam$^{-1}$~km~s$^{-1}$ gain in emission is not enough to explain the observations. Increasing the photodesorption rates by two orders of magnitude does not yield an higher intensity, so photodesorption alone can not explain the N$_2$H$^+$ emission originating in the outer disk. 

Interestingly, the radius at which model and observations start to deviate ($\sim$60~AU) is equal to the radial extent of the millimeter grains \citep[see e.g.,][]{Andrews2012}. The absence of large grains in the outer disk, not accounted for in our model, may influence the temperature structure, such that thermal desorption becomes effective, as shown for CO by \citet{Cleeves2016}. An increase in CO and N$_2$ desorption may then cause an increase in N$_2$H$^+$ in the disk outer region.

Photodissociation turns out to be an important process in N$_2$H$^+$ chemistry, so a logical question to ask is whether N$_2$ photodissociation is responsible for the outer edge of the N$_2$H$^+$ emission. N$_2$ self-shielding is not effective until the N$_2$ column density becomes $\gtrsim10^{15}$~cm$^{-2}$ \citep{Li2013}, so although the N$_2$H$^+$ layer below the CO snow surface extends over the entire disk in most models (see Fig.~\ref{fig:N2H+_abundances}), the N$_2$H$^+$ abundance outside $\sim$~100~AU is two orders of magnitude lower for N$_2$ abundances $\lesssim 10^{-6}$. However, despite an N$_2$H$^+$ layer throughout the entire disk for N$_2 > 10^{-6}$, the outer radius of the emission coincides with the outer boundary of the N$_2$H$^+$ surface layer, which is set by N$_2$ photodissociation. Only for N$_2$ abundances as high as $10^{-4}$ does the N$_2$H$^+$ emission extend over the entire disk. For lower abundances is the emission thus truncated due to N$_2$ photodissociation at the outer edge in this particular model.


\section{Discussion} \label{sec:Discussion}

To study the robustness of N$_2$H$^+$ as tracer of the CO snowline, we model the N$_2$H$^+$ distribution for the disk around TW Hya using a simple chemical model and simulate the resulting emission with the radiative transfer code LIME. The N$_2$H$^+$ column density peaks $\sim$5--30~AU outside of the CO snowline, for all physical and chemical conditions tested. Furthermore, the N$_2$H$^+$ emission peaks generally not at the same radius as the column density, and can be up to 53~AU offset from the CO snowline. Only for very low total, i.e. gas plus ice, CO abundances ($\sim$10$^{-7}$) can the emission peak inside the snowline, although the column density does not. Results for the different models are summarized in Table~\ref{tab:Results}. Fitting the N$_2$H$^+$ column density using a power law with the inner radius assumed to be at the CO snowline can thus generally only produce an outer boundary to the snowline location. 
 
Triggered by the question on how N$_2$H$^+$ can be abundant in protoplanetary disks in spite of very similar freeze-out temperatures for CO and N$_2$, \citet{Aikawa2015} performed a chemical model study of the N$_2$H$^+$ distribution. They attributed its presence to the conversion of CO to less volatile species. However, the models presented in this work predict an N$_2$H$^+$ layer even for canonical CO abundances of $\sim10^{-4}$. Nonetheless, the conclusions that the absolute abundances of CO and N$_2$ are important and the N$_2$H$^+$ abundance can peak at a temperature below the CO and N$_2$ freeze-out temperature are reinforced by our models for many different CO and N$_2$ abundances. Results on the effect of the CO and N$_2$ binding energies and cosmic ray ionization rate are also in good agreement.

\begin{figure*}
\centering
\includegraphics[width=17cm,trim={0 0cm 0cm 0.5cm},clip]{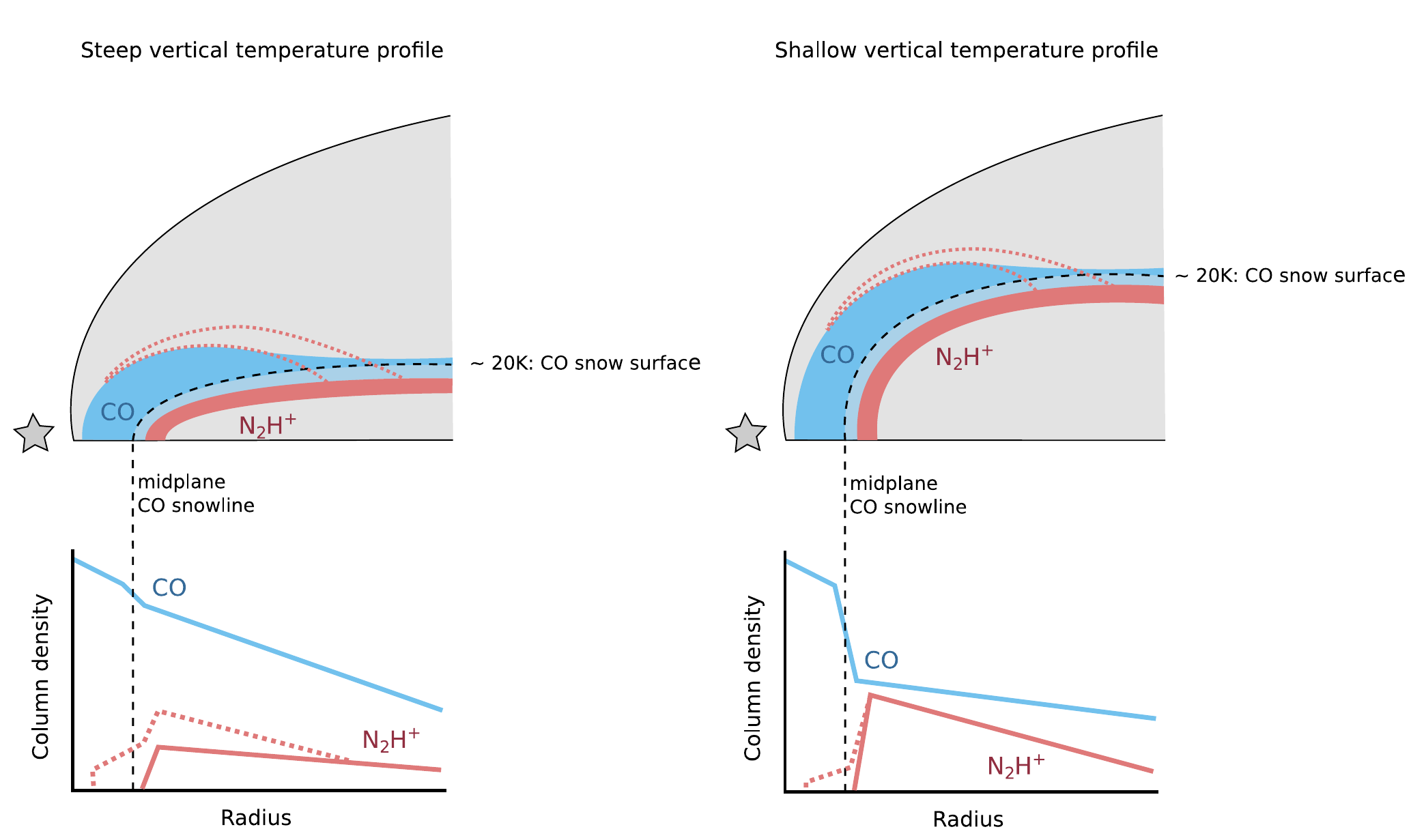}
\caption{Schematic representation of the distribution of gas-phase CO (blue) and N$_2$H$^+$ (red) in disks with either a steep vertical temperature profile, as for TW Hya (left), or a shallow vertical temperature profile (right). These differences can be due to different degrees of grain settling. To highlight the region around the CO snowline, the vertical direction depicts scale height, $z/r$. The dashed black contour represents the CO snow surface and the light blue area directly outside this contour shows that, at the snow surface, the gas phase abundance drops by 50\%. The N$_2$H$^+$ surface layer is indicated by dotted red lines. The predicted column density profiles are shown below. For N$_2$H$^+$, the column density profile is shown with (dotted line) and without (solid line) the surface layer. The vertical dashed black line indicates the location of the midplane CO snowline. } 
\label{fig:Discussion}
\end{figure*}

\citet{Aikawa2015} also report the presence of N$_2$H$^+$ in layers higher up in the disk in their full chemical model (in line with \citealt{Walsh2010} and \citealt{Cleeves2014}), but they do not perform a radiative transfer calculation to explore whether this contributes significantly to the resulting emission. Here we show that the discrepancy between column density and emission maxima is caused by such a surface layer that is present in models with CO less than $\sim$5 times as abundant as N$_2$, due to a small difference in the CO and N$_2$ photodissociation rates. Although CO is more than an order of magnitude more abundant than N$_2$ in the ISM, CO can be underabundant in the gas-phase in protoplanetary disks. This underabundance used to be attributed to photodissociation and freeze-out \citep{Dutrey1997,vanZadelhoff2001}, but recent studies concerning in particular TW Hya, suggest that on top of these well-known processes, CO is also depleted in the warm molecular layer and indeed inside the snowline \citep{Favre2013,Nomura2016,Schwarz2016}. Moreover, observations of [C I] lines indicate a general carbon depletion in this disk \citep{Kama2016}. The results presented here show that N$_2$H$^+$ is very sensitive to the gas-phase CO abundance, and the best fit to the observed emission is acquired for a total CO abundance of $3\times10^{-6}$, consistent with the CO depletion scenario. To achieve such a low CO gas-phase abundance in the models, a low total CO abundance is required, as the amount of CO present in the gas phase depends on the available ice reservoir. This suggests that CO frozen out in the outer disk may be trapped in the ice or converted to more complex species. Other possiblilities are that it has been locked up due to grain growth on the way to the inner disk or locked up in even larger bodies like planetesimals. The overprediction of N$_2$H$^+$ emission inside the CO snowline as compared to the observations may indicate that some of the CO is trapped in other ices with higher binding energies, such as CO$_2$ and H$_2$O, since this will allow gradual release of additional CO when these species desorb off the grains at higher temperatures. 

The contribution of the surface layer to the total N$_2$H$^+$ emission seems to depend on the disk physical structure. In the TW Hya model, the high degree of dust settling results in a steep vertical temperature gradient. This confines the CO snow surface, and hence the associated N$_2$H$^+$ layer, close to the midplane. For a less settled disk, the vertical temperature gradient is shallower and the N$_2$H$^+$ layer resides higher up in the disk. The N$_2$H$^+$ column density just outside the CO snowline is much higher in the latter case and therefore the contribution from the N$_2$H$^+$ surface layer is significantly lower. This is further aided by the lower gas density in the surface layer due to its increased scale height (see Fig.~\ref{fig:Discussion}). For CO abundances $\gtrsim10^{-6}$ the N$_2$H$^+$ emission then traces the column density, while the emission is shifted to larger radii in models with substantial grain settling. 

Differences in disk vertical structure may also help explain why the CO snowline can be observed directly with CO isotopologues in some disks, but not in others. The higher up in the disk the CO snow surface resides, the larger the CO column density decrease across the snowline, simply because the CO-depleted region extends to larger heights (see Fig.~\ref{fig:Discussion}). This may explain why in TW Hya no sharp drop in CO column density is seen around the snowline, on top of the global CO depletion \citep{Nomura2016,Schwarz2016}. The rise in column density inward of 10 AU may be the result of release of trapped CO at the CO$_2$ and H$_2$O snowlines. On the other hand, in HD 163296, both C\element[][18]O and N$_2$H$^+$ emission can be reproduced by a sharp change in column density at roughly the same radius \citep{Qi2015}. The fitted CO freeze-out temperature occurs, for the physical model adopted for HD 163296 by these authors, at a radius of 85--90~AU, while the N$_2$H$^+$ emission can be reproduced by a column density profile with inner radius between 84 and 98 AU. These results are consistent with the results shown here that the N$_2$H$^+$ distribution peaks outside the CO snowline. The better agreement between CO and N$_2$H$^+$ emission could mean that the CO snow surface is located higher up in the disk. As the HD 163296 disk is inclined with respect to the line of sight, \citep[e.g.,][]{Dominik2003}, this hypothesis could be tested by deriving the height at which the N$_2$H$^+$ layer resides from channel maps. Another possibility is that there is no N$_2$H$^+$ surface layer due to the much stronger UV radiation field of the Herbig Ae star HD 163296 as compared to the \mbox{T Tauri} star TW Hya, so that the N$_2$H$^+$ emission follows the column density. In addition, a strong drop in CO abundance may be easier to detect in a disk with a low global carbon and CO depletion.

The relationship between N$_2$H$^+$ and the CO snowline is thus more complicated than direct coincidence and a snowline location can generally not be derived from only a power law fit to the observed N$_2$H$^+$ emission. For disks with the CO snow surface high above the midplane, e.g. due to a low degree of grain settling, the N$_2$H$^+$ emission seems to generally trace the column density peak quite well. The then obtained outer boundary for the snowline can be improved if a CO column density profile can be derived from C\element[][18]O observations. On the other hand, when the N$_2$H$^+$ emission is dominated by a surface layer, e.g. in a very settled disk, chemical modeling is required. If the CO snow surface is close to the midplane, the CO column density change across the snowline will be small and C\element[][18]O observations will be less helpful (see Fig.~\ref{fig:Discussion}). Detailed knowledge about the disk vertical physical structure are thus required to translate N$_2$H$^+$ emission into a CO snowline location. Comparing emission from multiple N$_2$H$^+$ transitions can provide information on to what extent the emission is dominated by an N$_2$H$^+$ surface layer, and thus how well it traces the column density. Higher spatial resolution may also reveal significant contribution from a surface layer, as multiple components may be concealed in a broad emission peak at low resolution.


\section{Conclusions}  \label{sec:Conclusions}

In this work, we modeled the N$_2$H$^+$ distribution and resulting emission for the disk around TW Hya using a simple chemical network. Our main conclusions regarding the robustness of N$_2$H$^+$ as a tracer of the CO snowline are listed below. 

\begin{enumerate}
\item For the adopted physical structure and binding energies, freeze-out and thermal desorption predict a CO snowline at 19 AU, corresponding to a CO midplane freeze-out temperature of 20 K. This is smaller than inferred by \citet{Qi2013}. 

\item A simple chemical model predicts the N$_2$H$^+$ column density to peak at least $\sim$5~AU outside the CO snowline for all physical and chemical conditions tested. This offset shows an increasing trend with CO abundance, suggesting that the N$_2$H$^+$ distribution is dictated by the amount of CO present in the gas phase, rather than its snowline location.

\item In addition to the N$_2$H$^+$ layer outside the CO snow surface, N$_2$H$^+$ is  predicted to be abundant in a surface layer where the gas-phase N$_2$ abundance exceeds that of CO due to a small difference in the photodissociation rates. Only in models with N$_2$/CO $\lesssim$ 0.2 is no surface layer present. 

\item The contribution of this surface layer to the total N$_2$H$^+$ emission depends on the disk vertical structure. For the adopted physical structure for TW Hya, in which the large grains have settled toward the midplane, the simulated N$_2$H$^+$ emission is dominated by the surface layer. This causes the emission to shift to even larger radii, up to $\sim$50~AU beyond the snowline. The influence of the surface layer is much smaller in a less settled disk, and in this case the N$_2$H$^+$ emission does roughly trace the column density. 

\item The extent of the N$_2$H$^+$ surface layer, and therefore the shift of the emission peak, also depends on the CO abundance. Moreover, the peak integrated intensity depends on the N$_2$/CO ratio. Together, this suggests that N$_2$H$^+$ may help constrain CO and N$_2$ abundances in protoplanetary disks, provided a representative model of the physical structure is derivable from existing observations. 

\item An N$_2$H$^+$ distribution based on the freeze-out and desorption balance for CO and N$_2$, and thus peaking directly at the CO snowline, produces an emission peak 7~AU closer to the star than observed. To reproduce the observed emission peak with the simple chemical model, a CO and N$_2$ abundance of $3\times10^{-6}$ is required. This is in agreement with a global CO and carbon depletion in TW Hya. The N$_2$H$^+$ surface layer predicted by the simple chemical model is necessary to fit both the location and the intensity of N$_2$H$^+$ emission peak. 

\item The cosmic ray ionization rate influences both the N$_2$H$^+$ intensity as well as the positions of the column density and emission maxima, while only the peak positions change with different CO and N$_2$ binding energies. 

\item Underprediction of the emission from the region depleted of millimeter grains (radii larger than $\sim$60~AU) reinforces the idea that N$_2$H$^+$ may be very sensitive to the physical structure of the disk. 
\end{enumerate}

The relationship between the N$_2$H$^+$ distribution and the CO snowline location is thus more complicated than initially assumed and simple parametrized column density fits provide only upper boundaries for the snowline radius. Instead, more detailed modeling is needed to derive the CO snowline location from N$_2$H$^+$ emission, and as shown in this work, a simple chemical model seems to be sufficient. However, detailed knowledge of the disk physical structure is required. On the other hand, the sensitivity to CO and N$_2$ abundance and physical structure suggests that N$_2$H$^+$ may be a more versatile probe, capable of constraining CO and N$_2$ abundances, and the thermal structure of protoplanetary disks.


\begin{acknowledgements} 
We thank Michiel Hogerheijde for sharing the reduced image cube of the ALMA N$_2$H$^+$ $J$ = 4--3 observations in TW Hya, Ilse Cleeves for fruitful discussions and the anonymous referee for useful comments to improve the manuscript. Astrochemistry in Leiden is supported by the European Union A-ERC grant 291141 CHEMPLAN, by the Netherlands Research School for Astronomy (NOVA) and by a Royal Netherlands Academy of Arts and Sciences (KNAW) professor prize. M.L.R.H acknowledges support from a Huygens fellowship from Leiden University. C.W. acknowledges support from the Netherlands Organisation for Scientific Research (NWO, program number 639.041.335). This paper makes use of the following ALMA data: ADS/JAO.ALMA\#2011.0.00340.S. ALMA is a partnership of ESO (representing its member states), NSF (USA) and NINS (Japan), together with NRC (Canada), NSC and ASIAA (Taiwan), and KASI (Republic of Korea), in cooperation with the Republic of Chile. The Joint ALMA Observatory is operated by ESO, AUI/NRAO and NAOJ.
\end{acknowledgements}


\bibliographystyle{aa} 
\bibliography{N2H+}


\begin{appendix}


\section{Chemical model} \label{ap:chemmodel}

\subsection{Freeze-out and desorption balance}

The balance between freeze-out onto dust grains and desorption back into the gas phase can be written as: 

\begin{equation}
  k_\mathrm{d} n_\mathrm{s}(\mathrm{CO}) = k_\mathrm{f} n_\mathrm{g}(\mathrm{CO}),
\end{equation} 
where $n_\mathrm{s}$(CO) is the CO ice abundance, $n_\mathrm{g}$(CO) the CO gas abundance, and $k_\mathrm{f}$ and $k_\mathrm{d}$ are the freeze-out and desorption rates, respectively. For N$_2$ a similar equation holds. 

The freeze-out rate depends on the gas temperature $T_\mathrm{g}$ and is given by
\begin{equation}
  \label{eq:k_freezeout}
  k_\mathrm{f}= \left< v \right> \sigma_{\mathrm{grain}} n_{\mathrm{grain}} S \hspace{0.1cm} \mathrm{s}^{-1},
\end{equation}
where $\left< v \right> = \sqrt{8k_\mathrm{B}T_\mathrm{g}/{\pi}m}$ is the mean thermal velocity of molecules with mass $m$ in the gas phase at a temperature $T_\mathrm{g}$, $k_\mathrm{B}$ is Boltzmann's constant, $\sigma_{\mathrm{grain}}n_{\mathrm{grain}}$ is the average dust-grain cross section per unit volume and $S$ is the sticking coefficient \citep{Allen1977}. The sticking coefficients for CO and N$_2$ are taken to be 0.9 and 0.85, respectively, which are the lower limits found by \citet{Bisschop2006}. Assuming them to be unity does not significantly affect the results. The average dust-grain cross section per unit volume can be written as 
\begin{equation}
\sigma_{\mathrm{grain}}n_{\mathrm{grain}} = C \int_{a_{\rm{min}}}^{a_{\rm{max}}} a^{-3.5} \left(\pi a^2\right) da \hspace{0.2cm} \mathrm{cm^2 \, cm}^{-3},
\end{equation}
where $a$ is the grain radius, and $a_{\mathrm{min}}$ and $a_{\mathrm{max}}$ are the minimum and maximum radius, respectively, for the assumed grain size distribution. The constant of proportionality, $C$, can be derived from the total dust mass per unit volume, $\rho_{\mathrm{dust}}$:  
\begin{equation}
\rho_{\rm{dust}} = C \int_{a_{\rm{min}}}^{a_{\rm{max}}} a^{-3.5} \rho_{\rm{bulk}} \left(\frac{4\pi}{3} a^3\right) da \hspace{0.2cm} \mathrm{g \, cm}^{-3},
\end{equation} 
where $\rho_{\rm{bulk}}$ is the bulk density of the dust grains. For the adopted grain size distributions, this yields results similar to assuming a typical grain size of 0.1 $\mu$m, because, although a significant fraction of the grains have grown to larger sizes, the small grains still provide the bulk of the surface area.  

Thermal desorption will be considered here as the only desorption process, which is appropriate for volatile molecules such as CO and N$_2$. The desorption rate then depends on the dust temperature $T_\mathrm{d}$ \citep{Allen1977} and can be written as 
\begin{equation}
  \label{eq:k_desorption}
  k_\mathrm{d} = \nu_0 \exp \left( - \frac{E_\mathrm{b}}{T_\mathrm{d}} \right) \hspace{0.1cm} \mathrm{s}^{-1},
\end{equation} 
where $E_\mathrm{b}$ is the binding energy of a species to the dust grain and $\nu_0$ is the characteristic vibrational frequency of an adsorbed species in its potential well \citep{Allen1977},
\begin{equation}
  \nu_0 = \sqrt{\frac{2n_\mathrm{s}E_\mathrm{b}}{\pi^2m}} \hspace{0.1cm} \mathrm{s}^{-1},
\end{equation}
where $E_\mathrm{b}$ is again the binding energy and $n_\mathrm{s} = 1.5 \times 10^{15}$ cm$^{-2}$ is the number density of surface sites on each dust grain \citep{Hasegawa1992}. For CO and N$_2$ a binding energy of 855~K and 800~K \citep{Bisschop2006} are adopted, respectively, in the fiducial model.

\subsection{Simple chemical network}

The rate coefficients for the two-body reactions in the simple chemical network are given by the standard Arrhenius equation:
\begin{equation} \label{eq:ratecoeff}
  k = \alpha \left( \frac{T_{\rm{g}}}{300} \right) ^{\beta} \exp \left( -\frac{\gamma}{T_{\rm{g}}} \right) \hspace{0.1cm} \mathrm{cm}^3 \, \mathrm{s}^{-1},
\end{equation}
where $T_{\rm{g}}$ is the gas temperature, and $\alpha$, $\beta$ and $\gamma$ can be found in the UMIST database for Astrochemistry \citep{McElroy2013} (see Table~\ref{tab:RateCoefficients}), while the cosmic-ray ionization rate of H$_2$ is taken to be:
\begin{equation}
  \zeta = 1.2 \times 10^{-17} \hspace{0.1cm} \mathrm{s}^{-1}
\end{equation}
in the fiducial model \citep{Cravens1978}. The disk surface density is not high enough to shield cosmic rays, i.e. \mbox{$<$ 96 g cm$^{-2}$} everywhere, so no attenuation takes place toward the disk midplane. 

The photodissociation rates of CO and N$_2$ can be written as
\begin{equation} \label{eq:pd_rate}
  k(r,z) = k_0(r,z) \, \mathrm{exp}(-\tau_{\mathrm{UV}}) \, \Theta[N(\mathrm{H}_2),N(\mathrm{X})] \hspace{0.2cm} \mathrm{s}^{-1},
\end{equation}
where $k_0$ is the unattenuated photodissociation rate, $\mathrm{exp}(-\tau_{\mathrm{UV}})$ is a dust extinction term and $\Theta[N(\mathrm{H}_2),N(\mathrm{X})]$ is the shielding function for shielding by H$_2$ and self-shielding \citep{vanDishoeck1988}. For CO and N$_2$, photodissociation occurs through line absorption and the unattenuated photodissocation rate are therefore given by
\begin{equation} 
  k_0(r,z) = \sum_{\lambda} \frac{\pi e^2}{mc^2}\lambda^2 f \eta I(\lambda) \hspace{0.2cm} \mathrm{s}^{-1},
\end{equation} 
where $f$ is the oscillator strength for absorption from the lower to the upper level, $\eta$ is the dissociation efficiency of the upper level, $I(\lambda)$ is the mean intensity of the radiation field in photons cm$^{-2}$ s$^{-1} \AA^{-1}$ at wavelength $\lambda$, and $\frac{\pi e^2}{mc^2}$ is a numerical factor \citep{vanDishoeck1988}. For the stellar radiation field, the attenuation by dust can be calculated from the attenuated $F_{\mathrm{att}}^*(r,z)$ and unattenuated radiation field $F_{\mathrm{unatt}}^*(r,z)$:
\begin{equation}
  \mathrm{exp}(-\tau_{\mathrm{UV}}^{*}) = \frac{F_{\mathrm{att}}^*(r,z)}{F_{\mathrm{unatt}}^*(r,z)} \hspace{0.2cm} \mathrm{photons} \, \mathrm{cm}^{-2} \mathrm{s}^{-1} \AA^{-1}, 
\end{equation}
while the attenuation of the interstellar radiation field can be derived from the visual extinction:
\begin{equation}
  \tau_{UV}^{ISRF} = \gamma A_V,
\end{equation}
where $A_V$ is the optical depth in magnitudes and the factor $\gamma$ depends on the dust properties. The shielding functions depend on the H$_2$ and CO or N$_2$ column densities. As photons from the star can reach a molecule either through the inner edge of the disk or from above or below, an effective column density is used assuming $N_\mathrm{H} = 1.59 \times 10^{21} A_V$. This is translated into effective column densities for CO and N$_2$ by scaling with the respective abundances. Interstellar radiation on the other hand can only penetrate from the outside of the disk and for simplicity the radial contribution is ignored as this will provide only a minor contribution to the radiation field in the disk. Therefore, for shielding against the interstellar radiation field vertical column densities are used. For the adopted models, photodissociation is dominated by the stellar radiation field.

\subsection{Photodesorption}

Although for volatile species such as CO and N$_2$, thermal desorption is the main desorption mechanism, the dust density in the outer disk may be low enough for UV photons to penetrate to the disk midplane, such that photodesorption may become effective. Therefore, photodesorption is included in the chemical model when studying the outer edge of the N$_2$H$^+$ emission (model CH-PD). The photodesorption rate for species X is given by: 
\begin{equation} \label{eq:photodesorption}
k_\mathrm{PD} = F_\mathrm{UV}(r,z) Y \sigma_{\rm{grain}}n_\mathrm{grain} \frac{n_\mathrm{s}(\mathrm{X})}{n_\mathrm{ice}} X_\mathrm{M},
\end{equation}
where $F_\mathrm{UV}(r,z)$ is the UV photon flux at position $(r,z)$, $Y$ is the number of molecules desorbed per incident photon, $\sigma_\mathrm{grain}n_\mathrm{grain}$ is again the average dust-grain cross section per unit volume and $n_\mathrm{s}(\mathrm{X})/n_\mathrm{ice}$ is the fraction of species X in the ice. The term $X_\mathrm{M}$ accounts for the fact that only molecules in a few monolayers near the ice surface ($M_\mathrm{surf}$) can desorb \citep{Andersson2008}, and can be written as 
\begin{equation}
X_\mathrm{M} = \frac{M_\mathrm{total}}{M_\mathrm{surf}}, 
\end{equation}
where $M_\mathrm{total}$ is the total number of monolayers, i.e. the number of molecules in the ice divided by the number of available binding sites on the dust grains. Assuming only molecules in the top four monolayers can desorb ($M_\mathrm{surf} = 4$), $X_\mathrm{M}$ is set to 1 for $M_\mathrm{total} \geq 4$. For mixed ices with an equal amount of CO and N$_2$, the yield for CO and N$_2$ is similar and we adopt values of $1.4 \times 10^{-3}$ and $2.1 \times 10^{-3}$ molecules photon$^{-1}$, respectively \citep{Bertin2013,Paardekooper2016}. Since the uncertainty in photodesorption rates is high \citep[see e.g.][]{Paardekooper2016}, also the extreme case with a yield 100 times higher is considered. 


\section{N$_2$H$^+$ $J$=4-3 peak integrated intensity} \label{ap:intensity}

\begin{figure}
\centering
\includegraphics[trim={0 16.5cm 0cm 1.3cm},clip]{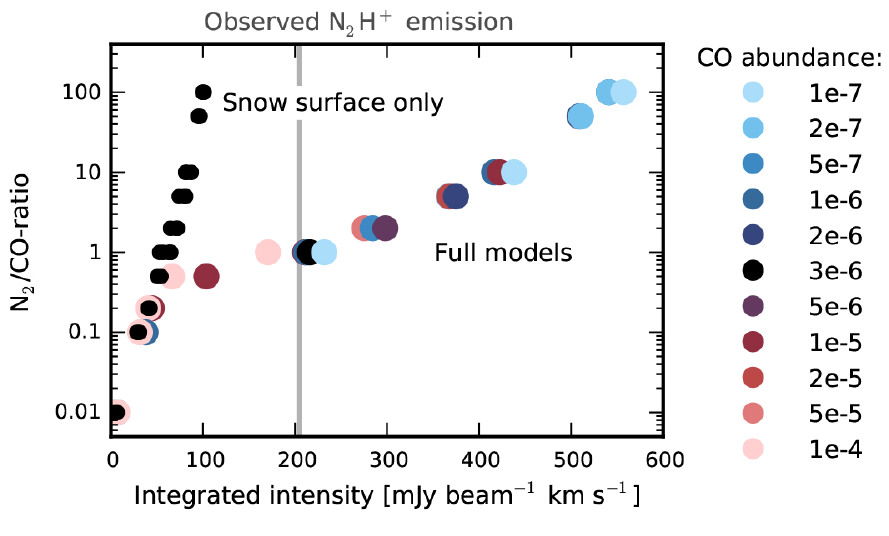}
\caption{N$_2$H$^+$ $J$=4--3 peak integrated intensity for models with different CO and N$_2$ abundances (models CH). The simulated emission is convolved with a $0\farcs63\times0\farcs59$ beam. Black circles represent snow surface only models, i.e. N$_2$H$^+$ removed above the CO snow surface. The color of the circles for the full models represent the CO abundance. Note that some points may overlap. The grey line indicates the observed intensity.} 
\label{fig:EmissionIntensity}
\end{figure}

\begin{figure}
\centering
\includegraphics[trim={0 17.3cm 0cm 1.5cm},clip]{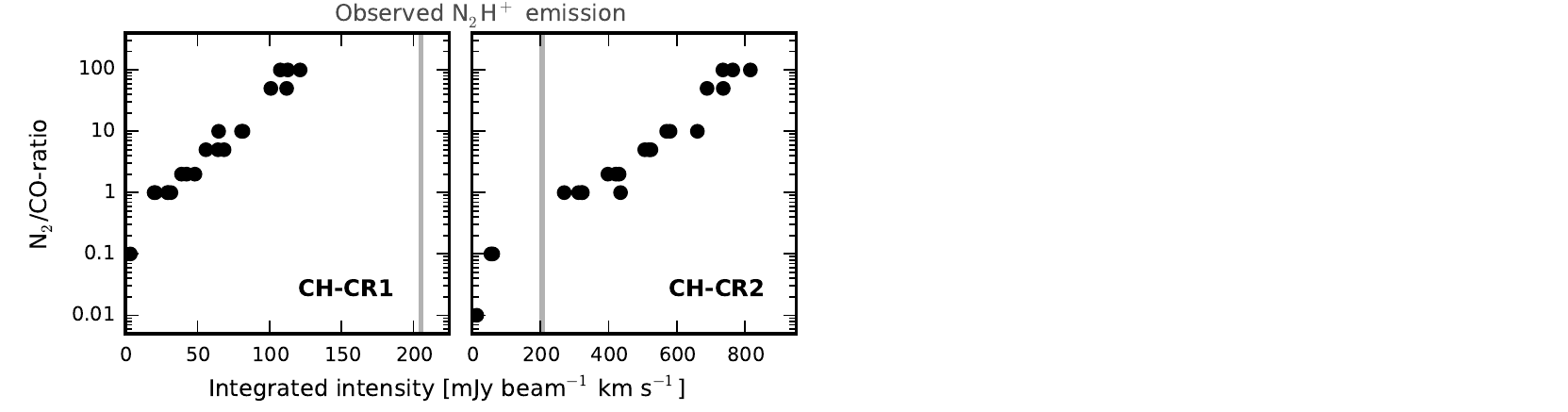}
\caption{N$_2$H$^+$ $J$=4--3 peak integrated intensity as function of N$_2$/CO ratio for models with a cosmic ray ionization rate of $\zeta = 1\times10^{-19}$ s$^{-1}$ (model CH-CR1; left panel) or $\zeta = 5\times10^{-17}$ s$^{-1}$ (model CH-CR2; right panel). The simulated emission is convolved with a $0\farcs63\times0\farcs59$ beam. The grey line indicates the observed intensity.} 
\label{fig:Intensity_ZetaCR}
\end{figure}

In Figs.~\ref{fig:EmissionIntensity}~and~\ref{fig:Intensity_ZetaCR}, we show the simulated N$_2$H$^+$ $J$=4-3 peak integrated intensity for model CH and the models with different cosmic ray ionization rates (models CH-CR1 and CH-CR2), respectively.

%

\section{N$_2$H$^+$ line ratios} \label{ap:Lineratios}

\begin{figure*}
\centering
\includegraphics[width=17cm,trim={0 10cm 0cm 1cm},clip]{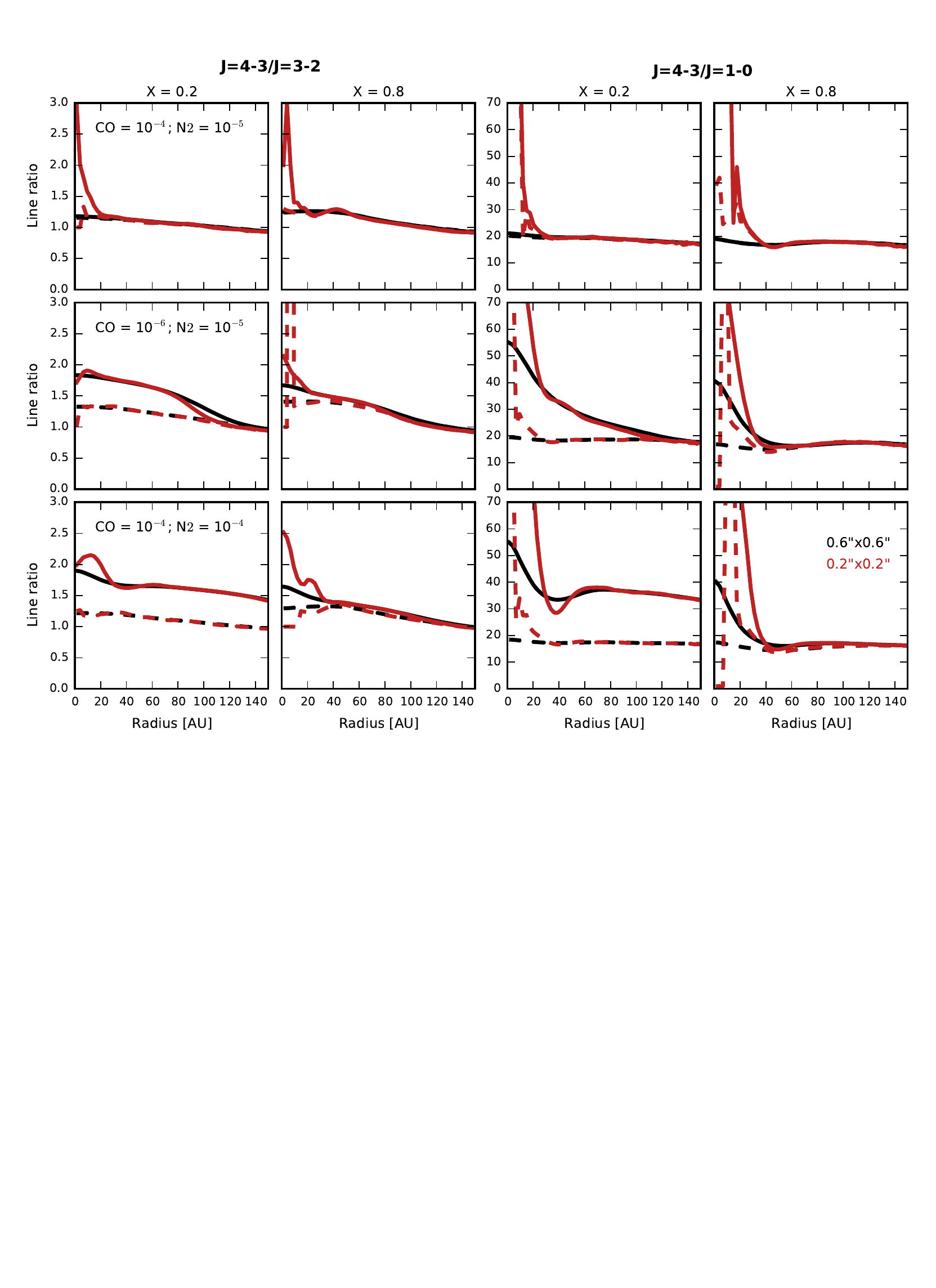}
\caption{N$_2$H$^+$ $J$=4--3/$J$=3--2 (left panels) and $J$=4--3/$J$=1--0 line ratios (right panels) for three different CO and N$_2$ abundances, as indicated in the leftmost panels, in models with large grains settled to 20\% (column one and three) or 80\% of the small grain scale height (column two and four). Dashed lines show the snow surface only models, while solid lines represent the full models. The simulated emission is either convolved with a $0\farcs63\times0\farcs59$ (black lines) or $0\farcs2\times0\farcs2$ beam (red lines).} 
\label{fig:Lineratios}
\end{figure*}

N$_2$H$^+$ $J$=4--3/$J$=3--2 and $J$=4--3/$J$=1--0 ratios for model CH and model CH-$\chi$0.8 are presented in Fig.~\ref{fig:Lineratios}. The three depicted CO and N$_2$ abundances represent models with a large N$_2$H$^+$ surface layer (CO = $10^{-4}$, N$_2$ = $10^{-5}$, top row), a smaller N$_2$H$^+$ surface layer (CO = $10^{-6}$, N$_2$ = $10^{-5}$, middle row) and no N$_2$H$^+$ surface layer (CO = $10^{-4}$, N$_2$ = $10^{-4}$, bottom row), as shown in Fig.~\ref{fig:N2H+_abundances}.

\section{Photodesorption} \label{ap:Photodesorption}

\begin{figure*}
\centering
\includegraphics[width=17cm,trim={0 16.6cm 0cm 1.2cm},clip]{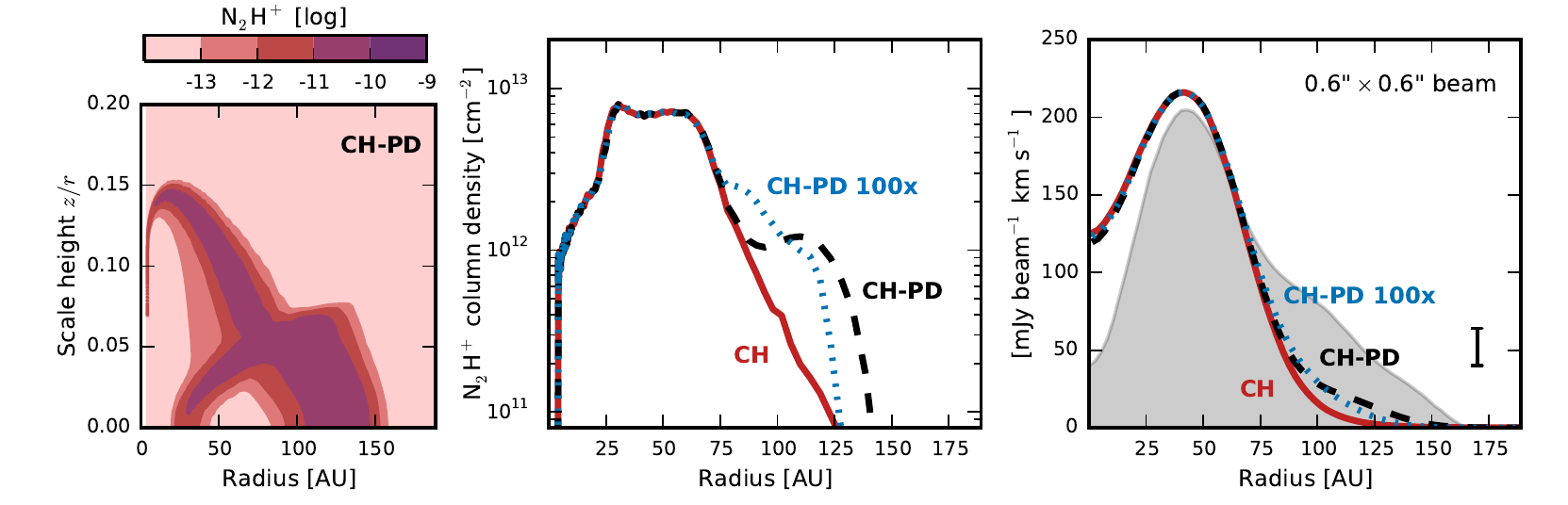}
\caption{N$_2$H$^+$ distribution (left panel), column density (middle panel) and radial $J$=4--3 integrated line intensity profile (right panel) when photodesorption is included in the chemistry (model CH-PD with CO and N$_2$ abundances of $3\times10^{-6}$; dashed black lines). The dotted blue lines show a model with photodesorption rates increased by a factor 100 (model CH-PD 100x), and the red solid lines show the fiducial model without photodesorption (model CH). The simulated emission is convolved with a $0\farcs63\times0\farcs59$ beam. Observations by Q13 are shown in grey in the right panel with the 3$\sigma$-error depicted in the lower right corner.} 
\label{fig:Emission_Photodesorption}
\end{figure*}

The N$_2$H$^+$ distribution, column density profile and \mbox{$J$=4--3} integrated line intensity profile for the chemical model with photodesorption included (model CH-PD) are shown in Fig.~\ref{fig:Emission_Photodesorption}. 

\end{appendix}

\end{document}